\documentstyle[preprint,aps,epsfig]{revtex}
\begin{document}
\title{ Analysis of NN Amplitudes up to 2.5 GeV:\\
An Optical Model and Geometric Interpretation}
\author{H.V.~von~Geramb$^{a,b}$, K.A.~Amos$^a$, H.~Labes$^b$ 
and M.~Sander$^b$}
\address{$^a$ School of Physics, University of Melbourne, Parkville 3052,
Australia\\ $^b$ Theoretische Kernphysik, Universit\"at Hamburg,
Luruper Chaussee 149, D--22761 Hamburg}
\maketitle
\begin{abstract}
We analyse the SM97  partial wave amplitudes
for nucleon--nucleon (NN) scattering to 2.5 GeV, in which
 resonance and meson
 production effects are evident for energies above pion 
production threshold. 
Our analyses are based upon boson exchange or quantum 
inversion potentials
 with which the sub-threshold  data are fit perfectly. 
Above 300 MeV they are extrapolations, to which complex
short ranged  Gaussian potentials 
are added 
in the spirit of the optical models of nuclear physics and of
 diffraction models of high energy physics. 
The data to 2.5 GeV are all well fit. 
The energy dependences of these  Gaussians are very smooth 
save for precise effects caused by the known
$\Delta$ and N$^\star$ resonances. With this approach, we confirm that
the geometrical implications of the 
profile function found from diffraction  scattering are pertinent in the 
regime 300 MeV to 2.5 GeV and that the  overwhelming part of meson production
comes from the QCD sector of the nucleons
when they have a  separation of their centres of 
1 to 1.2 fm.
This analysis shows that the elastic NN scattering
data above 300 MeV can be understood with a local potential 
operator as well as has  the data below 300 MeV. 
\end{abstract}

\section{Introduction}
The nucleon--nucleon (NN) interaction at low and medium energy is a timely
 topic given the experimental efforts being made at institutions such as 
 IUCF, TRIUMF, SATURN, CELSIUS,
 and COSY. It is a timely topic also theoretically given 
the plethora of models of NN scattering in vogue. 
Below pion threshold (which we take to be synonymous with 300 MeV throughout
this paper),
phenomenology is rather simple as empirically
there is  only the
deuteron bound state,
the elastic scattering and 
NN$\gamma$ bremsstrahlung. 
For this domain
of energies below 300 MeV, there exist excellent experimental
data and
several potential models whose parametrisations give fits
with $\chi^2\sim 1$.  There have been many fine presentations of the 
experimental~\cite{arnd,byst87}
and  theoretical~\cite{myhr88,mach89}
developments for this energy regime also.

Above pion threshold, the experimental situation is
 excellent also~\cite{arnd}.
But as there are many inelastic channels,
 the available  experimental information is
less complete.
Nevertheless, in the energy regime 300 MeV to 1 GeV,
a number of experiments have produced data of such quality
that  existing models of NN scattering are severely tested.
The models predicated upon quality fits to NN 
scattering data below the pion threshold have to be 
modified if they are to be used as a starting point for
analyses of the higher energy data.
Notably they must be varied to account for the various meson 
production thresholds and also to account for effects of known 
resonance structures  in the NN system.
Of the latter, the $\Delta$ and N$^\ast$ are the relevant
entities for the energy range considered, and the
effects resulting from interference of their
associated  scattering
amplitudes with those of other possible scattering processes 
are very evident in the structures of the cross-section data
and spin observables.
Those effects are not severely localised in energy 
as the resonances have large widths for the decay.
Indeed, amplitudes for N$\Delta$,
NN$^\ast$, $\Delta\Delta$ among others are important and affect 
NN scattering at all energies in the range
from threshold to over 2 GeV.
Some of the studies of these problems
based upon  boson exchange models
give  qualitative if not 
quantitative descriptions of the 
situation~\cite{popp87,elst88,samm90,valc94}.
Lomon~\cite{lomo82} has studied
such resonance phenomena in a different way by using
a boundary condition model.

In the energy range above 5 GeV, the NN scattering system
is one of many overlapping resonances and many open reaction  channels.
A consequence is that diffraction models, such
as epitomised by Glauber or Regge theories~\cite{coll77}, explain
very well the  measured total and soft interaction NN scattering cross
sections from about 5 GeV to the highest experimental
energy~\cite{jack74,kamr84,donn96,matt94}.
As the energy decreases to circa 2 GeV, more specific treatment
of the scattering process is needed to explain observation.
An optical model approach by Neudatchin {\em et al.}~\cite{neud91}
so far has been the only
way reasonable agreement with 2 to 6 GeV data has been found.
In fact, that study covered the entire energy range below 6 GeV.
But there was little data for high precision phase shift
analyses available for use with their analysis 
and they did not seek fits to data that qualify also as high precision.
In fact, then essentially phase shift
values to 1 GeV only were known with some 
confidence~\cite{arnd}.
This situation has changed drastically in the intervening years.
Today Arndt {\it et al.}~\cite{arnd97} have investigated 
elastic scattering data for energies up to
2.5 GeV and they have defined partial wave 
scattering amplitudes which are available from SAID~\cite{said},
as are a wide range of other options.

A key feature in all studies of partial wave amplitudes
has been the attributes of the chosen phase shift specifications.
Until recently the data to 300 MeV led to diverse solutions
from various groups~\cite{arnd,bugg90,pwa93}.
Qualitatively they obtain the same results to 300  MeV with
the exception of the $^1P_1$ channel and the mixing angle  $\epsilon_1$.
Inclusion of extra data sets (to 1 GeV) in an extension of the 
method of analysis~\cite{arnd,byst87,bugg90}
helped resolve some of that ambiguity. New data then extended the range
of the analysis to 1.6 GeV~\cite{arnd}, from which a confident
solution for the amplitudes was defined to about 1.2 GeV.
Finally, the data from COSY pushed that limit to 2.5 GeV
with a confidence interval to about 1.75 GeV~\cite{arnd97}.
A  noted result of this most recent development
is that the NN partial wave amplitudes 
are particularly smooth functions of energy
allowing for large width resonance structure associated
with $\Delta$ (1232 MeV) and N$^\ast$ (1440 MeV) formation.
These two resonances have relatively small contributions
to the overall amplitudes, but the precision of analysis
is such as to establish them. The remaining essentially
smooth behaviour of the scattering amplitudes was  not anticipated.
There were expectations that dibaryon resonances
effects would exist as well~\cite{jaff79,star91,lomo82}.
If such do exist in the range to 1.8 GeV, then either they must
have very small coupling strengths or they must have 
very narrow or extremely large widths.

The character of the scattering amplitudes up to 2.5 GeV
is consistent with the optical potential concept. Thus
we suggest a potential approach to the analyses of that 
data based upon extrapolating a high precision
NN interaction, established by its fit to data below 
300 MeV, to energies above that and correcting
with energy dependent optical potentials in each 
partial wave. By so doing, we account for the spin, isospin,
and momentum dependences of underlying boson
exchange mechanisms. At the highest energy,
as so many partial wave amplitudes contribute
to scattering, the optical potential
scheme should simplify and ultimately
merge to the optical disc of diffraction models.
A consequence of this approach is 
a geometric picture of NN scattering from the
highest energies down; a picture which has been correlated 
to the Regge theory with pomeron
exchanges~\cite{jack74,kamr84,donn96,matt94}.

Our theoretical efforts to analyse NN data to 2.5 GeV,  begins either with 
boson exchange models, in particular the nonlinear
 one solitary boson exchange potential OSBEP, or quantum inversion.
 Observed data in the sub-threshold region $<$300 MeV
are reproduced perfectly by those two very different approaches with
 \ OSBEP defining a potential in 
momentum space while inversion leads to a local coordinate space
one~\cite{jaed97,kohl94}. As we indicated above, 
use of these potentials as well as of the Paris~\cite{laco80},  
Nijmegen~\cite{stok93}, AV18~\cite{av18} and  
Bonn--CD~\cite{mach96}
 potentials for energies above 300 MeV are
extrapolations. They all give similar results and could be used as a real
background potential in an extended application to account
 for meson exchanges for energies $>$300 MeV.
For  energies above 300 MeV, our model
is to  add to any of  the background potentials, a
real and imaginary potential with Gaussian form factors
whose parameters are adjusted 
to give  fits to the SM97 partial wave amplitudes~\cite{arnd97}. 
Smooth energy dependent results have been found 
that are consistent with the structures in the 
SM97 data, which indicate resonances in several partial
waves, notably the $P_{33}$(1232) and $P_{11}$(1440),
 on an otherwise smooth energy dependent background.
The optical potentials are complex and short ranged
typically of nucleon size  that is known from analyses 
of electron scattering off a nucleon.
This implies the first of our conjectures that
production processes are localised 
at and within  the confinement surface of a nucleon.
The results we display also supports our second conjecture that
the geometry of the profile function, known from high 
energy diffraction scattering, remains valid 
at lower energies and especially
in the $P_{33}$ and $P_{11}$ resonance dominated region.
It is this result that lead us to expect also
that meson production is a unique  QCD aspect applicable from 
threshold (300 MeV) up to highest energies. 

In  Sect.~II, the basis and form of the optical potential we
investigate for NN scattering above threshold is defined. The results 
of our calculations then are given in Sect.~III and the geometric picture we 
associate with them is presented in Sect.~IV. Finally a summary of this work
and the conclusions we have drawn from our results are presented in Sect.~V.

\section{Optical potential -- its basis and form}
 
To describe NN scattering, we adopt a coordinate space view.
At very long range, electromagnetic
interactions alone are important,
but as the range decreases, boson exchange attributes
become increasingly effective. The onset is at about 15 fm
with the exchange of a pion. As the range shortens further,
then $\sigma$, $\rho$, $\omega$ meson and baryon exchanges add in.
At the shortest distances, inside 0.8 fm typically,
all NN potentials have strong repulsion.
The precise character of this core is not a sensitive quantity,
so far as low to medium energy NN scattering is concerned.
Past success of use of soft and hard core potentials 
reflects that lack of sensitivity.

The boson exchange models which we ascribe to the medium
and long range attributes, 
are developed in momentum space.
The associated interactions are nonlocal.
With reasonable values for the meson-nucleon coupling constants
and form factor cut-offs,
these boson exchange models  give quality fits to the data
that lead to their nomination as high precision 
interactions~\cite{mach89,stok93,av18,mach96,jaed97}.
That is also the case with other 
approaches such as those with explicitly momentum dependent
potential models~\cite{laco80,stok93}, energy independent
partial wave potential 
models~\cite{kohl94,stok93,av18}, and, with somewhat different approaches, 
the Moscow potential model~\cite{mosc84},
and the MIT boundary condition model~\cite{lomo82}.
These approaches are motivated  differently in their
formulation but in the end all  give essentially the same
on-shell NN $t$-matrices below threshold.

We are particularly interested in those
potentials obtained by use of inverse scattering
theories that are predicated upon a Schr\"{o}dinger equation
as the equation of motion.
This is an ill-posed problem since only
discrete data with 
uncertainties in the finite interval
0 to 300 MeV are input.
Solution of the inverse problem then
requires an interpolation and an extrapolation
of the data.  We constrain that extrapolation
so that the $S$-matrix remains unitary
at all energies.  Below the pion threshold this is a very good
approximation since bremsstrahlung is the only open 
channel and, as that has a small cross section,
it is customary to neglect this violation of
unitarity.
The problem then is well-posed and, by
using Gel'fand-Levitan-Marchenko  
equations, real and energy independent
inversion potentials have been constructed
partial wave by partial wave~\cite{kohl94,sand96,sand97}. 
By dint of this construction the on--shell $t$-matrices 
(0 to 300 MeV) are perfectly reproduced.

In recent years, attempts have been made to discern between
these diverse model views
by seeking explicit effects in data due to the
off--shell properties
of the associated NN $t$-matrices.
 Studies of three nucleon systems,
of bremsstrahlung, and in microscopic nucleon--nucleus
optical models are examples.
So far no study has been able to discriminate
one model form over any other or even set a preference
order. 
While all of the potential models considered 
are relevant physically only for the range 0 to 300 MeV,
mathematically there is no prohibition in obtaining
solutions for energies above threshold.
The extrapolations are shown for several of these
potentials in Figs.~1--4.
Qualitatively they are similar
in all channels save for those in which the 
known resonances have big effects;
notably in  ${}^1D_2$,  ${}^3F_3$ and ${}^3PF_2$ channels.
All of these potentials are purely real so that
they result in unitary $S$-matrices, as they do not incorporate
production or annihilation of mesons;
effects which are important in analyses of data
above 300 MeV. 
This is evident in Figs.~\ref{sm97ppeta} and \ref{sm97npeta}
in which the
absorption, $\eta= \vert S \vert$, is shown for
 uncoupled  proton-proton
and neutron-proton channels respectively.

There exist extensions to boson exchange models which
incorporate particle production explicitly~\cite{popp87,elst88}.
They reproduce well observed  NN and NN$\pi$
data up to 1 GeV. At the time  data above that energy
were sparse.
Even so, these calculations are extremely complex;
much more so than for the boson exchange  models that
are their base. Also the number of adjustable
parameters involved increase markedly
with every additional element in the theory.  
Most seriously from our point of view, however,
is that the conventional boson exchange amplitudes are varied
from the forms optimal below 300 MeV.
Consequently the NN potentials are affected at all radii
and the meson production would not be as localised as we
believe it to be.
We conjecture  that meson production is a genuine QCD
effect and so, in a geometric view, 
emanate from the QCD bag. 
It is also the case that
the partial wave amplitudes are very
smooth functions of energy,
giving credence to our view that a model with far fewer
degrees of freedom should suffice.
In light of the above,
we seek a simpler
phenomenological approach to interpret the
elastic scattering and reaction cross sections
above 300 MeV.  It is the optical model.
Use of complex optical potentials to analyse hadron-hadron scattering is
not new~\cite{jack74}. Most studies also have shared the general 
characteristics of that optical potential by its links to the strong
absorption model that works so well with high energy scattering data. For
NN scattering to 6 GeV, such an 
approach has been used recently as well~\cite{neud91}.
But there now is quite excellent data to 2.5 GeV and there are diverse basic
NN interactions that give high precision fits to sub-threshold data for use as sensible background interactions..

The optical model for scattering is a concept that is well
developed in
nuclear physics from both a purely phenomenological view
as well as from a microscopic (folding model) one.
That is especially the case for nucleon scattering
from nuclei with projectile energies to 400 MeV and more.
The phenomenological approach was developed first as 
a means to categorise much data and the smooth behaviour
with energy, target mass and projectile type of those nuclear
optical potentials
indicated a sensibility of the model and gross properties of
nuclear systems which more fundamental approaches should 
encompass.
The
microscopic models of nucleon-nucleus scattering
were developed subsequently.
With them
excellent results 
can now be obtained whether the approach is based on a model
in momentum 
space~\cite{arel96} or on one in coordinate space~\cite{amos97}.
The  complex optical potentials
predict  nucleon-nucleus scattering that agree very well 
with measured cross sections and spin observables for
all nuclei between $^3$He and $^{238}$U.  These proton-nucleus optical model
results correlate with intrinsic 
nuclear structure  consistent with
electron scattering form factors from those nuclei.

It may be argued that an optical model approach
for study of NN scattering is not necessary
as the extended boson  exchange models will provide
the essential information that a QCD based theory must emulate.
For all the reasons listed above, this is not our opinion. 
Our use of an optical model approach to the analysis of NN 
scattering above pion threshold is predicated in part
 upon the successful use of 
that approach to proton-nucleus scattering analyses
but also because of the 
folding to get the proton--nucleus optical potentials 
is similar in spirit to what has been proposed for  quarks
by Nachtmann {\it et al.}~\cite{land87}.
Also there is a synergy of optical 
potential methods between low energy and high energy
scattering studies and 
we seek its form for NN scattering over the entire 
energy range. The criterion that we 
have a sensible result will be that of a smooth
 behaviour of the properties 
of the potentials found and a consistent geometric
interpretation of what the complex potentials reflect.
We comment on this later for first we show that analyses made using a
relativistic Schr\"odinger equation are pertinent.

\newcommand{\fdag}{\not\kern-.25em}

It is generally accepted that a valid covariant description
of NN scattering formally is given
by the Bethe--Salpeter equation
\begin{equation} \label{BeSal}
{\cal M} = {\cal V} + {\cal V}{\cal G}{\cal M}\ ,
\end{equation}
where  $\cal M$ are invariant amplitudes
that are based upon irreducible
diagrams as contained in $\cal V$, and 
$\cal G$ is a relativistic propagator.
This equation serves generally as an ansatz for
approximations. Of those, the three dimensional reductions
are of great use and, of those,
the Blankenbecler and Sugar~\cite{blan66} reduction
gives an equation that has received most attention for 
applications with NN scattering~\cite{part70,mach89}. 
In this approach
an effective potential operator is introduced which one
identifies as the
NN interaction potential.
This reduction is obtained from the
integral equation  (\ref{BeSal}), which 
in terms of four-momenta~\cite{itzy80} is
\begin{equation}
{\cal M} ( q^\prime,q;P ) = {\cal V} ( q^\prime,q;P ) + 
\int d^4k\ {\cal V} ( q^\prime,k;P )\  {\cal G} (k;P)\ {\cal M} ( k,q;P)\ ,
\end{equation}
with the propagator
\begin{equation}
{\cal G} (k;P) = { i \over (2 \pi)^4} \left[ { \frac12\fdag P
+ \fdag k + M \over (\frac12 P + k)^2 - M^2 + i \varepsilon}\right]^{(1)}
\left[ { \frac12\fdag P
+ \fdag k + M \over (\frac12 P + k)^2 - M^2 + i \varepsilon}\right]^{(2)}\ .
\end{equation}
The superscripts
refer to the nucleon (1) and (2) respectively, and in 
the CM system, $P = (\sqrt{s},0)$,
 with  total energy $\sqrt{s}$. 
The Blankenbecler-Sugar reduction of the propagator $\cal G$
is to  use the covariant form
\begin{equation}
{\cal G}_{\rm BbS} (k,s) = - {\delta (k_0) \over (2 \pi)^3 } { M^2 \over E_k}
{\Lambda_+^{(1)} ({\bf k}) \Lambda_+^{(2)} (-{\bf k}) \over
\frac14s - E_k^2 + i \varepsilon}\ ,
\end{equation}
where the positive energy projector is given as
\begin{equation}
\Lambda_+^{(i)} ({\bf k}) = \left( { \gamma^0 E_k - \vec{\gamma} \cdot
{\bf k} + M \over 2M} \right)^{(i)}.
\end{equation}
Then 
the three-dimensional equation,
\begin{equation}
{\cal M} ({\bf q}^\prime,{\bf q}) = {\cal V} ({\bf q}^\prime,{\bf q}) +
\int { d^3k \over (2 \pi)^3 }{\cal V} ({\bf q}^\prime,{\bf k})
 { M^2 \over E_k} {\Lambda_+^{(1)} ({\bf k}) \Lambda_+^{(2)} (-{\bf k}) \over
{\bf q}^2 - {\bf k}^2 + i \varepsilon}
{\cal M} ({\bf k},{\bf q})\ ,
\end{equation}
is obtained. Taking matrix elements with 
only positive energy spinors, 
an equation with  minimum relativity is obtained
for the  NN $t$-matrix, namely
\begin{equation}
{\cal T} ({\bf q}^\prime,{\bf q}) =  {\cal V} ({\bf q}^\prime,{\bf q}) +
\int { d^3k \over (2 \pi)^3}  {\cal V} ({\bf q}^\prime,{\bf k})
 { M^2 \over E_k} {1 \over {\bf q}^2 - {\bf k}^2 + i \varepsilon}
{\cal T} ({\bf k},{\bf q})\ .
\end{equation}
With the substitutions
\begin{equation}
 T ({\bf q}^\prime,{\bf q})
= \left( M \over E_{q^\prime} \right)^{\frac12} 
{\cal T} ({\bf q}^\prime,{\bf q})
\left( M \over E_{q} \right)^{\frac12}
\end{equation}
and
\begin{equation} \label{emtrans}
V ({\bf q}^\prime,{\bf q})
= \left( M \over E_{q^\prime} \right)^{\frac12}
{\cal V} ({\bf q}^\prime,{\bf q})
\left( M \over E_{q} \right)^{\frac12},
\end{equation}
we obtain an expression equivalent to the non-relativistic  
 Lippmann--Schwinger equation, 
\begin{equation}
 T ({\bf q}^\prime,{\bf q}) =  V ({\bf q}^\prime,{\bf q}) +
\int { d^3k \over (2 \pi)^3}  V ({\bf q}^\prime,{\bf k})
 {M \over {\bf q}^2 - {\bf k}^2 + i \varepsilon}
 T ({\bf k},{\bf q})\ .
\end{equation}
This is  equivalent to the Schr\"odinger equation in
coordinate space, 
\begin{equation} \label{relschroe}
\left[-\Delta +M\, V({\bf r})-k^2\right] \psi({\bf r},  k)=0\ ,
\end{equation}
where $M$ is the reduced mass,
\begin{equation}
M={2\mu\over \hbar^2}={2\over \hbar^2}{m_1m_2\over m_1+m_2}\ .
\end{equation}
Because of  the $(M/E)$ factors in the transformation 
(\ref{emtrans}),
an explicitly energy independent potential, $V({\bf q'},{\bf q})$,
becomes
an energy dependent one, ${\cal V}({\bf q',q})$.
We note that
a proper relativistic wave equation 
would contain coupling to negative
energy solutions also,  but this we neglect. 
In the Schr\"odinger 
equation, (\ref{relschroe}), $k^2$ should be 
calculated relativistically, so defining the
 relativistic Schr\"odinger equation
which we have solved using an interaction of the form,
\begin{equation} 
V({\bf  r}) \to V_{NN}+V_{OMP}(r,s)+iW_{OMP}(r,s)
+{e^2Z_1Z_2\over r}\ ,
\label{SEeq}
\end{equation}
where $V_{NN}$ is
an energy independent background potential 
and ($V_{OMP},W_{OMP}$) is an energy dependent, complex
optical potential,
\begin{equation} V_{OMP}(r,s)=V_0(s)\exp{-r^2/a^2},\end{equation} and
\begin{equation} W_{OMP}(r,s)=W_0(s)\exp{-r^2/b^2}\ .
\end{equation}
Our choice of Gaussian form factors for the optical potential is based in part
upon the success of the Chou-Yang model~\cite{jack74} which shows that the
charge form factor of the proton determines the momentum transfers in their
approach.
The proton charge form factor is very well represented by a Gaussian and
folding two Gaussians yields again a Gaussian.
For coupled channels,  $V({\bf r})$ in 
 (\ref{SEeq}) becomes a 2$\times$2 matrix.
The optical potential search 
in this case is ambiguous 
if one uses the full matrix form, since the optical
model then has to account for flux losses into the production
channels as well as flux interchange between the coupled channels
themselves. 
The best situation would be to suppress the possibility
of flux interchanges between the explicitly included
coupled channels but such is not feasible within a search.
Thus we have estimated the optical potentials in coupled channel
 cases by using a two step procedure.
We run the coupled channel search twice
with the optical potential
 matrix restricted to act in channel ${}^3S_1$ (${}^3P_2$)
and ${}^3D_1$ (${}^3F_2$) respectively.
The search criteria then were solely the diagonal S matrix
elements of each channel in turn.

These equations are solved using partial wave expansions 
and so any of the coordinate 
space potentials
could be used as the background,
partial wave by partial wave.
We use the inversion potentials  since 
the inverse scattering approach 
always maps the latest phase shifts
as accurately as one wishes and 
permits a controlled extrapolation above 300 MeV.

To complete the specifications of our solutions
of the relativistic Schr\"odinger equations, we give
the relevant kinematics.
With $m_1$ being
the projectile and $m_2$ the target nucleon, 
 the 
 Mandelstam variable 
$s$, and the invariant mass $M_{12}$,
are given by
\begin{equation}
s=M_{12}^2=(m_1+m_2)^2+2m_2T_{\mbox{Lab}}=
\left(\sqrt{k^2+m_1^2}+ \sqrt{k^2+m_2^2}\right)^2\ ,
\end{equation}
while the relative momentum in CM system is
\begin{equation}
k^2 = { \displaystyle m_2^2 ( T_{Lab}^2 + 2 m_1 T_{Lab}) \over
\displaystyle ( m_1 + m_2 )^2 + 2 m_2 T_{Lab}}\ ,
\end{equation} 
which, for equal masses, reduces to
\begin{equation}
k^2 = \frac14 s- m_1^2\ .
\end{equation}
Integration of the partial wave components
of (\ref{relschroe}) is achieved using the Numerov
method to ascertain the asymptotic forms of the 
scattering solutions 
from which we get the phase shifts.

\section{Optical model analyses}

First we consider the background potentials we have used 
in our optical model approach~\cite{sand96}.
Given our primary interest in a geometric view of the scattering process,
 we seek background potentials that encompass the basic boson exchange
 processes as exactly as possible. This we define by virtue of a high
 precision fit to scattering data below threshold. In this manner we
 presuppose that the NN interaction at separation radii in excess of
 1 to 2 fm are established for any energy.
They are the potentials from inversion of
SM94 continuous fit phase shifts
of which we selected to show the $pp$ and $np$ T=1 channels in
Figs.~\ref{sm94scppt1phase} and \ref{sm94scnpt1phase}.
 The inversion potentials
 are  given in 
Figs.~\ref{sm94sct1pot}, \ref{sm94sct0pot} and \ref{sm94cct0t1pot} 
wherein are shown the potentials of neutron-proton (solid) and
proton-proton (dashed)  uncoupled and coupled  channels. The potentials
reproduce the continuous phase shift functions in every partial wave to
better than 0.02 degrees, which reflects our numerical accuracy used. The
continuous energy solutions have no error bars. The single channel T=1
phase shifts computed from inversion potentials fit
perfectly the 0 to 300 MeV energy region and its extrapolation 
to 1.6 GeV agree quite well with the real parts of the Arndt phase 
shifts~\cite{arnd}.
Inherently, the extrapolation is  given by the rational representation
of the data which form the input to inversion with the implication that
all phase functions are real and asymptotically 
decay $\lim_{k\to\infty}\delta(k)\sim o(1/k)$. This implies that the short
range interaction is either attractive, for phase functions which are
positive and remain positive at high energy, or repulsive, 
for phase functions which are negative and remain negative at high energy.
This choice of extrapolation permits evaluation of  singular
potentials with a behaviour near the origin $\sim 1/r$, and which imply
soft core potentials.
We have regularly updated our inversion potentials and used as input
the phase shift solutions PWA93~\cite{pwa93},
VV40, VZ40, FA91, SM94, SM95 as well as several other solutions from 
SAID~\cite{kohl94,sand96,sand97}. Any of these could have been used as
our background, although we consider the principal set to be PWA93
 from Nijmegen~\cite{pwa93}, 
SM94 and VZ40 from Arndt {\it et al.}~\cite{arnd}.
Only PWA93 single channel results are shown  in Fig.~\ref{pwa93scpot} as 
solutions to VZ40 are very similar to the potentials
found using SM94 and which have been shown before.
Qualitatively, the two sets of inversion potentials 
have the same structure but quantitatively they differ
 especially in the repulsive region.
These differences reflect the uncertainties in 
the extrapolation of phase shifts to higher energies.
The SM94 potentials have been chosen as background
because we have used the real parts of the SM94
phase shifts in the region 300 MeV to 1.6 GeV to
constrain the extrapolation.  Nijmegen 
phase shifts do not exist above 350 MeV. Nevertheless,
the high energy constraint is weak.
Now there are SM97 phase shift sets which
 extend to 2.5 GeV.  They are qualitatively
similar to the SM94 in the range 300 MeV to 1.6 GeV,
and as the actual phase shifts are
complex, we saw no fundamental reason to change the
extrapolation based upon the SM94 solution. While
the extrapolation determines how soft or hard
is the core,  the core radius is fixed largely
 from the low energy data. The core properties of the SM94 
proton-proton inversion  potentials
are displayed in Fig.~\ref{sm94t1core}.
Note that the core radii of the channels 
differ. Also, in all channels, the potential is repulsive
inside 0.8 fm. Of particular interest are the classical turning points
for our investigations of scattering to 2.5 GeV.
For the highest energy they are about 0.5 fm
increasing to about 1 fm at low energies.

With the inversion potentials as background, we 
used the optical potential approach to find high 
precision fits to the partial wave phase shifts
 (up to $L= 6$) and for energies to 2.5 GeV.
Guided by the Chou--Yang diffraction model~\cite{jack74,kamr84}, 
calculations have been made using Gaussian range 
values between 0.5 and 1.2 fm
which reflect the range of classical turning points
in the background potentials. We show only results where the real and
imaginary optical model potentials have the same range. The current analysis
show no evidence that they should differ.  
These values also span the radii of the little bag (0.5 fm) to the MIT bag
(1.2 fm).
The optical potentials strengths then were found to be smooth functions and
reproduce perfectly the continuous energy fit of SM97~\cite{arnd97}
 in the full energy range 0 to 1.6 GeV for neutron-proton scattering and
0 to 2.5 GeV for proton-proton scattering. 
These results are depicted by the solid lines in Figs. 1--6.
The results for all uncoupled channels are shown
in Figs.~\ref{07allsc} and \ref{10allsc}, and for coupled channels in 
Fig.~\ref{07allcc}. The real and imaginary potential strengths are shown 
in the left and right panels respectively. 
The results obtained from analyses of the SM97 data
 are portrayed by the open circles (pp data)
 and crosses (np data). 
The real and imaginary potential strengths are essentially charge independent.
Considering the real parts of these potentials first,
most channels have attractive real Gaussians
which shifts the net repulsions inward.
In contrast, up to
1.5 GeV, the ${}^1S_0$ and
the ${}^3P_0$ Gaussians are small 
but  add to the repulsive cores of the backgrounds.
The strong energy variation in the ${}^1D_2$ and ${}^3F_3$
channels reflect the $\Delta$ and N$^\ast$
resonance contributions to scattering; contributions that
had been predicted by
microscopic calculations~\cite{popp87}.
The large strengths in these channels simply reflects
centrifugal barrier shielding.
The imaginary parts of the potentials 
also show clearly the effects of the two known resonances.
These variations indicate central peak values of
$\sim$625  MeV in the ${}^1D_2$ and $\sim$900 MeV
in the ${}^3F_3$ channel.
As the $\Delta$ and N$^\ast$ have relative 
$L$ with the other nucleon of 0 and 1 respectively, 
the resonance strengths are distributed in many
partial waves so accounting for the variation in
effect that they have in the channels shown explicitly 
in these figures. This is well understood 
microscopically as well.
In the other uncoupled channels, with the exception of
the ${}^3P_0$, we observe a smooth imaginary Gaussian reflecting
an increased absorption with energy.
The ${}^3P_0$ case has a maximum absorption at $\sim$1 GeV
which corresponds to an invariant mass of $\sim$2.325 GeV.
At this point we note that the 
strong variation of absorption in the ${}^3P_0$
potential is associated with a $0^-;T=1$ state.
In addition,
analysis of neutron-proton data in the ${}^1P_1$
channel indicate a dramatic effect above 1 GeV.
Such could  be associated with a $1^-;T=0$ state. 
Despite the resonance features discussed above,
the optical potentials have very smooth strength variation
in all channels. They do not reflect any specific 
thresholds.
At the range chosen (0.7 fm for L=0--2, and 1 fm for L=3--6 ),
 all interaction strengths typically are  of several hundreds of MeV with 
fluctuations due to the prevailing resonances. An exception is the $^1P_1$
$np$ channel.
But this channel is sensitive to fine details in partial
wave analyses and is strongly correlated with the determination of the mixing
angle $\epsilon_1$. In some channels, 
and most clearly in the $^1S_0$,  $^3P_0$, $^3P_1$
 and $^1D_2$,   we notice that the potential
strengths have a kink at 1.75 GeV. However as
confidence in the phase shift analyses for energies above 1.6 GeV rest
solely on cross-section data without spin observables, 
no conclusions should be drawn about these structures as yet. We note again
that, for  higher  partial
waves, the interaction region determined from a Gaussian  is shielded 
from the centrifugal barrier and so   a much
larger strength is required
to achieve an effect. Such is evident in the results for the $^1G_4$,
$^3H_5$ and $^1I_6$ channels and at low energies.

We have studied the range dependence of the optical models in the 
interval  0.5 to 1.2 fm and show the results in three dimensional plots.
In  Figs.~\ref{potre} and \ref{potim} the real and imaginary strengths
are plotted as functions of kinetic energy and range 
and for the channels as indicated. Note that the real and imaginary potential
ranges were kept identical in these calculations.
Notably, as we expect with increasing range
the potential strengths decrease.
From these figures we note that, with
a channel independent range of 0.7 to 0.85 fm,
the optical models  have  evenly distributed
strengths in the dominating  L=0--3
channels,  while clearly maintaining  positions and widths of
the known resonances. This choice of  optical
model geometry with evenly distributed strengths of several hundred MeV
means that effective absorption (strengths from 0 to 50 MeV or more) occur
for radii larger than 0.7 fm. With Gaussian forms that absorption is quite 
localised and, from the history of optical model studies in general, we
infer that the maximum loss of flux in this case lies in a range 1 to 1.2 fm.

These conclusions have been drawn from analyses of data to 1.75 GeV.
We anticipate that such will remain the case
as sensitive data at higher  energies are gathered and analysed. We expect
that doing so with a
complex optical potential representation will result in a potential that
is less channel dependent until it merges with 
the diffraction models of high energy physics.
Thus, presuming phase shift analyses of new
 data in the 1 to 5 GeV range stay consistent
with the conjectures of smoothness,
there is a geometric connection of NN scattering at all energies.
The diffraction models are understood as the geometric
realisation of Regge theory with pomeron exchange.
No intrinsic structure of the nucleon
is identified from that data. Such requires
deep inelastic scattering studies.
A consequence of the continued geometric picture then is
that such intrinsic structures will not be
in evidence at low and medium  energy data
 save for the established roles of the $\Delta$ and N$^\star$ resonances.
This is a picture that is consistent also with the results obtained using
boundary condition models, like the P-matrix
formalism~\cite{jaff79}, and using the Moscow 
potential approach~\cite{mosc84}.

\section{The geometric picture}

The geometric picture we have of NN scattering can be 
divided in two segments; a soft and a hard part.
 The soft part we identify with the region 
outside $\sim$1 fm and in which the boson exchange
processes are the relevant mechanisms.  The
associated  potential strengths do not exceed 100 MeV
and are much less for most radii.
The hard part encompasses the internal region 
(inside 1 fm) and ultimately is QCD dominated.
The geometry of our optical model as well as of
high energy diffraction models place
production processes in the transition region of these two. 
However our view of `soft' is perhaps `super soft'
 in the high energy
terminology and our view of `hard' in that terminology may be `soft'.

At low energy, meson production is dominated by the 
processes involving intermediate resonance formation
of which the $\Delta$ resonance is the most important.
We consider just the $\Delta$ at this time. 
There are two extreme geometric pictures for its 
excitation.
These extremes are the result of 
potential model descriptions of $\pi$N scattering 
in the $P_{33}$ channel found using 
either  nonlocal (separable) interactions
in momentum space
or local interactions in coordinate space.

The first type of interaction is obtained  
by using the  separable potential from the Graz group~\cite{math85},
by using a  boson exchange model
 as has been done by Pearce and Jennings~\cite{pear91},
or by using  OSBEP~\cite{jaed98}.  Both boson exchange models include
the $\Delta$ as an s-channel resonance
whereby $\pi + N \to \Delta \to \pi + N$ is to be calculated. We
have used all three interactions.
The separable  Graz potential, for L=1 $\pi$N scattering, has the form 
\begin{equation}
V_1 (k, k^\prime) = g_1 (k) \, \lambda(s) \, g_1 (k^\prime)\ ,
\end{equation}
with the form factor
\begin{equation}
g_1 (k) = k {262.675  \over k^2 + (1.619)^2}\ .
\end{equation}
The parameter $\lambda(s)$
 in general is just a number independent of $s$ but   
in the resonant $P_{33}$ channel it was required to be
\begin{equation}
\lambda (s) = {1 \over s - m_0^2}
\end{equation}
with $m_0=1333.95$ MeV.

The second type of interaction is typified by
our inversion approach~\cite{sand97}. With these the scattering
can be interpreted as a t-channel
$\Delta$ exchange whereby $\pi + N \to N + \pi$.
The inversion result is a solution in coordinate
space and the wave functions we seek result directly with the method.
In contrast, the separable potential model and 
both boson exchange pictures were obtained by
solving the appropriate integral equations
in momentum space and then Fourier transforming 
into coordinate space. Thereby we obtained probability
distributions
in coordinate space for all interactions to allow geometric interpretation.
To support our claim that the two pictures are extreme,
we present in  Figs.~\ref{sepwf}, \ref{pjwf}, \ref{osbwf} and \ref{invwf} 
the moduli of those coordinate space wave functions 
in the $P_{33}$ $\pi$N channel  
as functions of the Mandelstam variable ($s$) 
in the regime of the $\Delta$ resonance.
 The radial distributions are very different.
The boson exchange results describe a molecular--like
system while the local inversion potential depicts
a highly concentrated $\pi N$ system where
the pion and nucleon are fused.
This has been a most astonishing result
as our initial expectations were that the two schemes
would infer that the $\Delta$ was an elementary excitation
of the nucleon interpretable as a reorientation
and alignment of valence quarks.  Since the {\em rms} radii
of both a pion and a nucleon are 
0.7 and 0.8 fm respectively,
the inversion picture implies that the $\Delta$
arises with practically a full overlap of the two hadrons
so that it would then
have a size of a nucleon and, concomitantly,
that the meson cloud
of a $\Delta$ is essentially that of the nucleon.
On the other hand, the results from the
boson exchange models suggest
that the $\Delta$ is far more extensive
implying that the meson cloud of a $\Delta$ is significantly
different to that of a nucleon. 
We have studied this situation also for
other resonances of hadronic systems and found in all cases such a
difference between the radial wave functions associated with
 separable and inversion potentials.

The results being such a surprise led us to look at
properties of other hadron-hadron scattering systems
for which phase shift analyses exist to allow inversion.
The inversion potentials for $\pi \pi$, $\pi$K, KN,
$\pi$N, and NN have been compared~\cite{sand96,sand97}
for cases where there exists low energy  resonances
in L=0--2 partial waves.
 These calculations revealed two  groups of short
 range potentials; one class being totally repulsive, the other
 having a barrier 
 inside of which is a strong attractive well.
 In Fig.~\ref{hadronpot} we display  this  geometry
 in a few cases, with which  resonances, $\Delta = \pi$N$(P_{33})$,
 $\sigma = \pi \pi (\delta_0^0)$, and
 $\rho = \pi \pi (\delta_1^1)$, 
  are associated. 
 We understand these potentials as effective operators
 which appropriately describe the dynamics of 
 the full system upon  projection into the elastic channel space.
In potential scattering terms then, the resonance is associated with
barrier penetration into an attractive well.
The $\alpha$-decays of heavy nuclei are classic examples of barrier penetration
in nuclear physics. The usual barrier for the $\alpha$-decay is broad and not 
high. In contrast, the $\Delta$ resonance is produced by one that is very thin
$\sim$0.1 fm, but extremely high $\sim$2--5 GeV. In both types of potentials,
the boundary conditions on the wave functions at the origin are that the
wave functions must vanish except for the L=0 case. The difference in 
establishing a resonance then lies with the matching of the internal
with external wave functions at the barrier. With the extremely high and
thin barrier, the dynamics of the internal system is practically
decoupled from the external one. Thus we associate no dynamics with the
thin barrier in contrast to the $\alpha$-decay situation in which the barrier
 is essential in the formation dynamics of the emerging $\alpha$.
 The $\pi$N system then comprises essentially two decoupled dynamic domains.
 Given the potentials we have found, this effective decoupling would hold to
 2--5 GeV, above which we anticipate the strong absorption model is valid.
 The $\Delta$ and N$^\star$ are L=1 resonances and are evident as such in
 cross sections since the L=1 wave function must vanish at the origin.
 For L=0 scattering the wave functions at the origin are not constrained 
and so no sharp resonance effect is likely to evolve. 
These considerations also imply that the resonances arise with
 practically a full overlap of the two hadrons, 
so that they too would then have a size of a single hadron.

There is then a consequence for NN scattering above threshold in
 that meson production is an emanation from the hard QCD (bag) region
 of one or the other
 nucleon, whether that be from non-resonant or resonant processes.

 We do not ascribe any further physical attribute to the potentials
 found. Rather they are just  effective local potential  operators
 that produce wave functions 
 at separations $\sim$1 fm consistent with boundary
 condition models~\cite{jaff79,lomo82}.
 Concomitantly, the data from which these results
 have been obtained
 then contain no further information on substructures of
 the systems. Such await advances in a QCD theory.

 We restate the surprising feature of the studies that while
 the boson exchange models are tuned also to produce similar appropriate
 boundary conditions, they do so at a significant larger radius
 $\sim$2 fm or more.  Differences like these
 mean that interpretation of results of
 momentum space calculations need be made carefully
 if those results are to be discussed from a geometric
  point of view.
 As a point in question, we can ask, `from where are the pions
 produced in NN scattering?'.
 From our local inversion model it is clear that such
 must come dominantly from the `hard' region or QCD sector.
 But it is not so clear that this viewpoint can be upheld
 with the boson exchange models, as they are used presently, without
baryon exchanges.

The formulation of a model, for example with NN scattering 
 the boson exchange model, relies on 
{\em a priori} assumptions which we associate with the physics of the
problem. The mathematical structure of
the specific boson exchange model formulations shows a
factorisation of terms. But experiment are compared with the full
product of amplitudes and results are not 
 very sensitive to the details of any isolated
process. This can be understood in terms of filter theory. To isolate
physics  uniquely
becomes very difficult in instances where many filters determine the
total result. Complicated models
may reproduce an  important
effect by a new or just  
 by  small modifications of the other existing components of the theory.
When that is so,  an 
implication may be that different models can  claim physical significance 
as they yield equally good fits to data. 
At present we have no practical and decisive experiment at hand which could
discern our t-channel view 
from the s-channel boson exchange model pictures despite the geometric
interpretations put forward. 

 \section{Summary and Conclusions}

 As the latest partial wave amplitude analyses
 of NN scattering data extend to 2.5 GeV and, notwithstanding
 known resonance effects, are very smooth functions of energy,
 the history of optical model approaches to data with
 such characteristics suggested to us that 
 we interpret NN scattering from 0 to 2.5 GeV
 in terms of a geometric model involving local 
 potential operators in each partial wave channel.

 Below threshold, local potential operators have been deduced 
 from high precision fits to the data. So also are
 momentum space models built upon boson exchange mechanisms.
 However, above threshold and with increasing energy, the
 semi--microscopic approaches using boson exchanges
 become very complex. The simplicity of
 an optical model approach with a complex potential
 to allow for pion production as flux loss from  the 
 elastic scattering channel commends itself
 as it is flexible in use and provides a connection between
 the low energy (boson exchange) regime and the high energy
 regime  where
 essentially a black disc absorption 
 replicates NN cross sections.
 Between the two energy regimes we place our optical model 
 and base it upon the low energy local potential operators
 as background.
 Any local case could be used as background for a model
 analysis of above threshold data.
 However, there are a number of reasons why we have used
 inversion potentials as the background
 in our optical model approach to NN scattering
 above threshold.
 First the inversion potentials are constructed so that
 high precision fits to partial wave phase
 shifts in the energy regime 0 to 300 MeV
  used as input in the inverse scattering 
 theory are retained.
 Second, in studies made using
 phase shifts chosen from a
 model calculated set, e.g. from the Bonn or Paris
 interactions, the inversion potentials found are
 consistent with the specific properties of those
 semi-microscopic  interactions as far as we can check.
 Third, semi--microscopic theories of NN scattering
 (0 to 300 MeV) give quality fits to phase shifts in most
 partial waves and so can be
 the underpinning description of physical processes
 for the inversion potentials.
 Fourth, 
 as more data has been gathered over the years,
 the results of phase shift analyses 
 vary in the precise values suggested
 for phase shifts in some channels, notably the ${}^3P_0$,
 and in the below threshold range (to 300  MeV) in
 particular.  Inverse scattering theory always maps the
 input and so has the flexibility to be tuned to ensure,
 as a background for analyses of data above 300 MeV 
 scattering, that the below threshold information
 currently in vogue will be exactly reproduced and
 maintained.
 Finally, 
 by using the inversion potentials as background,
 whatever one may glean from the character of
 optical potentials found by fits to above threshold
 data can be assured as due to underlying processes
 additional to those responsible for scattering 
 at   sub-threshold energies.

 The optical potentials we have found are consistent
 with properties of scattering known from
 other analyses.
 Specifically
 the geometries of absorption terms are consistent with
 the profile functions given by the diffraction models, and
 their energy variations 
 trace the properties
 of the known $\Delta$ and N$^\ast$ resonances.  
 Thus meson production, reflected in the extent of the imaginary 
 part of these optical potentials, would arise  effectively from 
 a `fused' system of the colliding hadrons,
 and the resonance would be an object of extent
 similar to a nucleon.
 The implication then is that meson production 
 would arise from almost complete overlap of the two colliding
  hadrons.
 This picture
 is consistent with what is obtained from  a local interaction
 of $\pi$N scattering
 at resonance energies.
 The associated wave functions imply that those resonances
 are local objects essentially the size of a nucleon.

 This view is consistent also with conclusions reached by Povh and 
Walcher~\cite{povh86} from their discussion of
elastic ${\bar p}p$ scattering.
They used an optical model approach to analyse cross-section data
identifying annihilation processes as flux loss associated with the 
imaginary part of that optical potential~\cite{dove83}.
 They deduce an absorption
probability in the S- and P-waves defined by 
\begin{equation}
p_\ell^{abs}(r)=W_{OMP}(r)\, u_\ell(kr)\, j_\ell(kr)\, r^2\ .
\end{equation}
These probabilities are quite sharply peaked functions, peaking at 
1--1.2 fm. Thus the absorption is quite localised. Interpreting their
results in terms of physical processes means that,
at very high energies, scattering is determined by
quark-quark interactions with
a range determined by the profile function. At lower energies that quark
interaction range manifests itself by the annihilation 
within the QCD sector of  the combined ${\bar p}p$ system. 

 In contrast, using a separable (momentum space) 
 model of NN scattering, 
 similar to those of boson exchange models,
 leads to a $\Delta$ resonance
 that has a molecule--like  probability distribution.
 The implication for pion production in NN scattering
 is that pions would be  released from long (spatial)
 ranged  attributes of the $\Delta$, to wit at least in part meson production
 would be `soft'.
 We do note however that scattering in higher partial waves
 deals essentially only with the periphery and so meson 
 production in those cases, if such be possible,
  may well be `soft' and involve 
 mesons from the meson cloud.

 The smooth behaviour of the optical potential strengths,
 the reflection in those variations of the known resonance 
 characteristics, and the consistency of the absorptive
 terms with the high energy profile functions, indicate
 that NN elastic scattering is not sensitive to any specific 
 QCD effect, save that such are necessary to specify
 intrinsic structures of the known resonances.
   All that seems needed to analyse the NN data is a 
 reasonable core radius and diffuseness  of the flux loss
 processes.
 We do note, however, that more 
 NN elastic scattering data are needed to pin down with 
 more certainty, the energy variations of the partial wave
 scattering amplitudes so that an even more discerning
 view may be taken about the specific optical potential
 characteristics. Also more data in the forward scattering
 region would be desired, i.e. for low momentum transfer 
  0.01--0.5 (GeV/c), as that would help confirm the link
 between our optical model and higher energy diffraction models.

 Finally, should perturbation calculations of small effects
 to scattering be of issue, 
 the optical model would be suitable to
 establish distorted waves in a distorted wave approximation
 analysis.     

\section*{Acknowledgement}
This work was supported in part by grants from the Forschungszentrum J\"ulich 
and the Australian Research Council. HVG and KAA acknowledge with gratitude
also the support of the University of Melbourne by means of a 
travel grant in aid.

\clearpage
\setlength{\unitlength}{1cm}
\begin{figure}[t]\centering
\begin{picture}(14,18)(0,3)
\epsfig{file=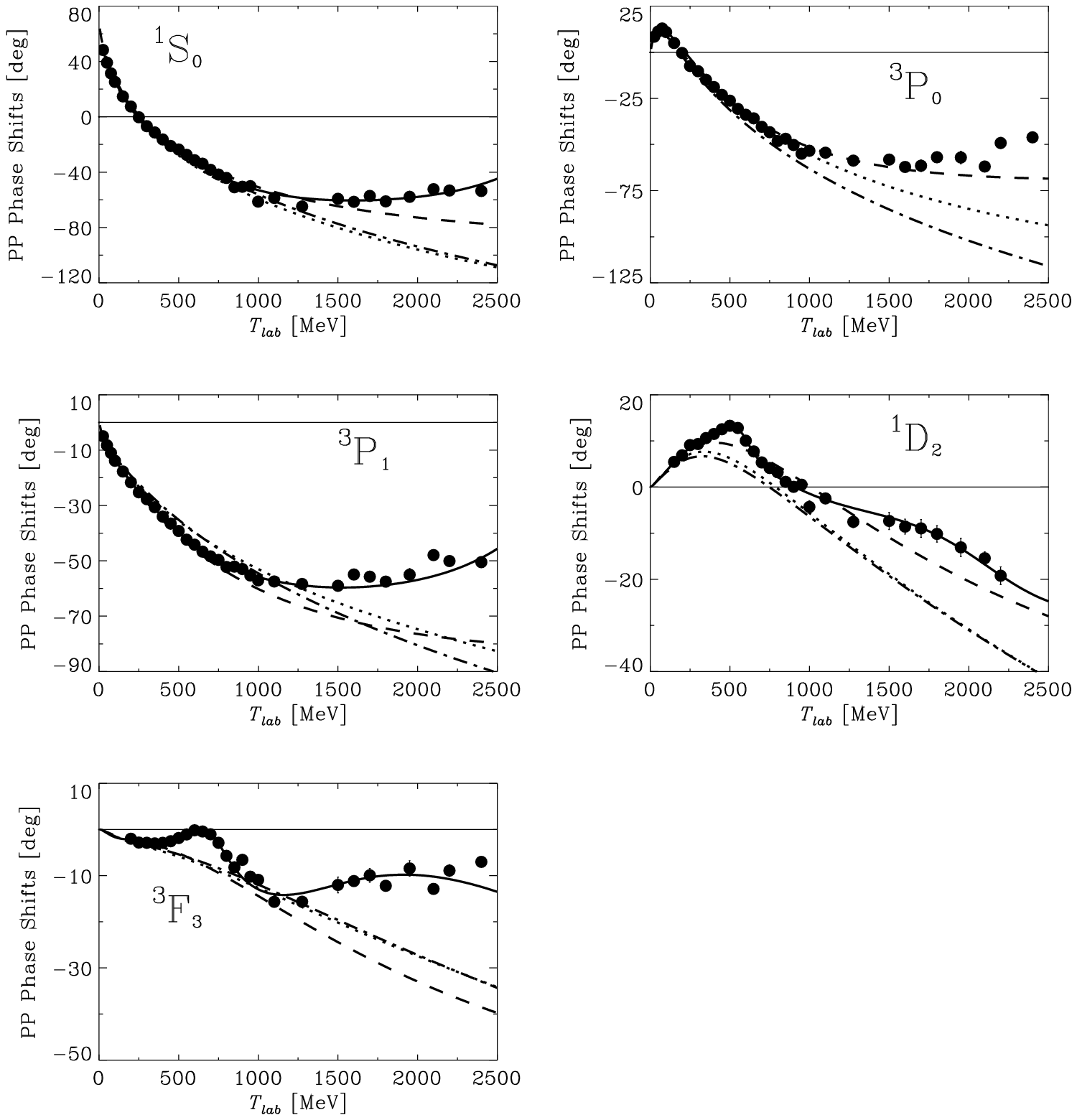,width=14cm}
\end{picture}
\caption[]{The phase shifts for
proton-proton scattering in uncoupled channels.
The full dots are  SM97 single energy fits while
the solid curves represent the SM97 continuous
energy fits as well as the final results of our optical model
searches. 
These are compared with 
the results of the inversion
potentials based upon SM94 values (dashed),
the OSBEP (dash-dotted),
the AV18 (long dashed) and the Bonn-B (dotted) results.
}
\label{sm97ppsc}
\end{figure}
\begin{figure}[t]\centering
\begin{picture}(14,20)(0,0)
\epsfig{file=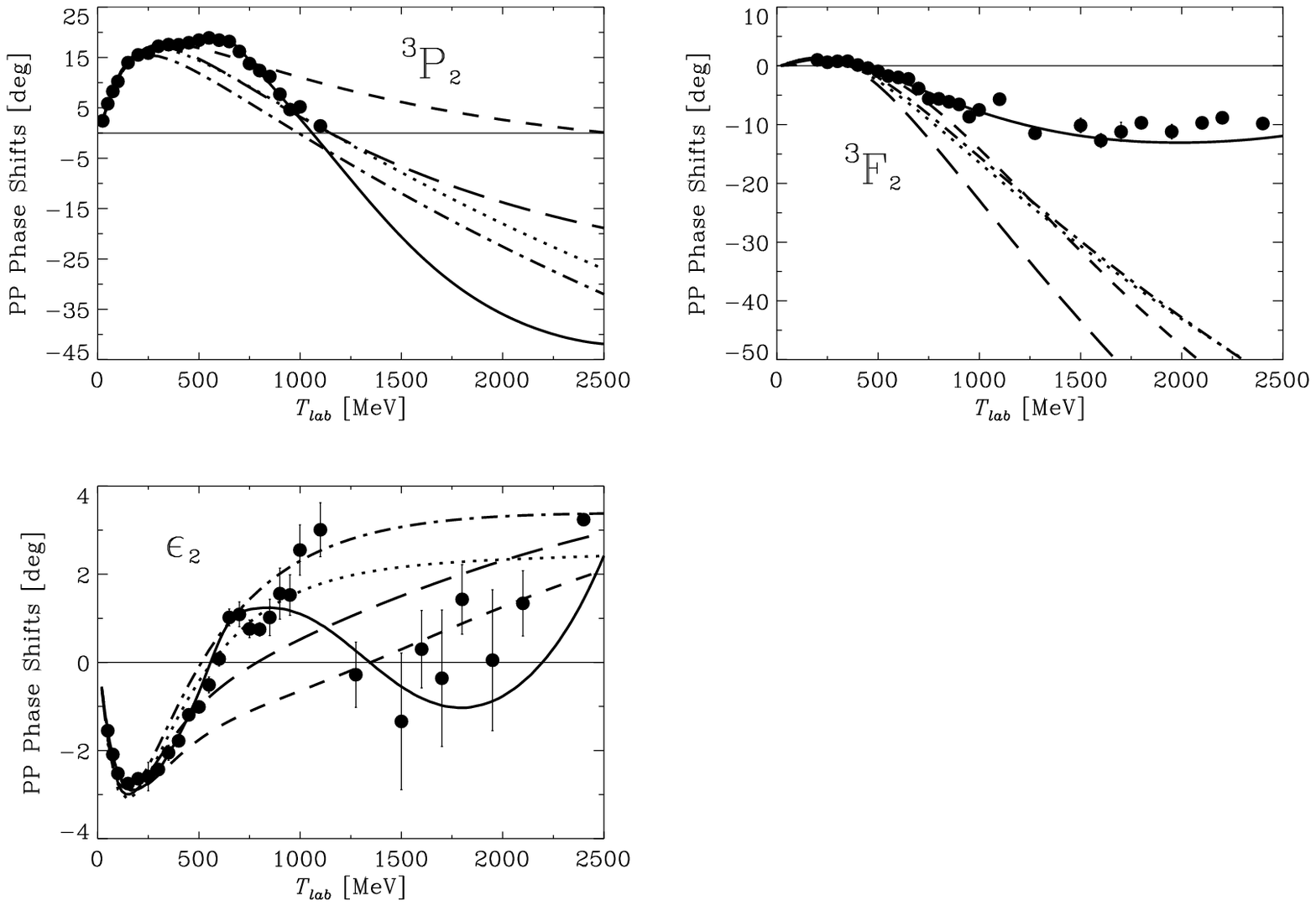,width=14cm}
\end{picture}
\caption[]{The phase shifts for
proton-proton scattering in 
the ${}^3PF_2$ coupled channels.
The nomenclature is as in Fig.~\ref{sm97ppsc}.}
\label{sm97ppcc}
\end{figure}
\clearpage
\begin{figure}[t]\centering
\begin{picture}(12,20)(0,-2)
\epsfig{file=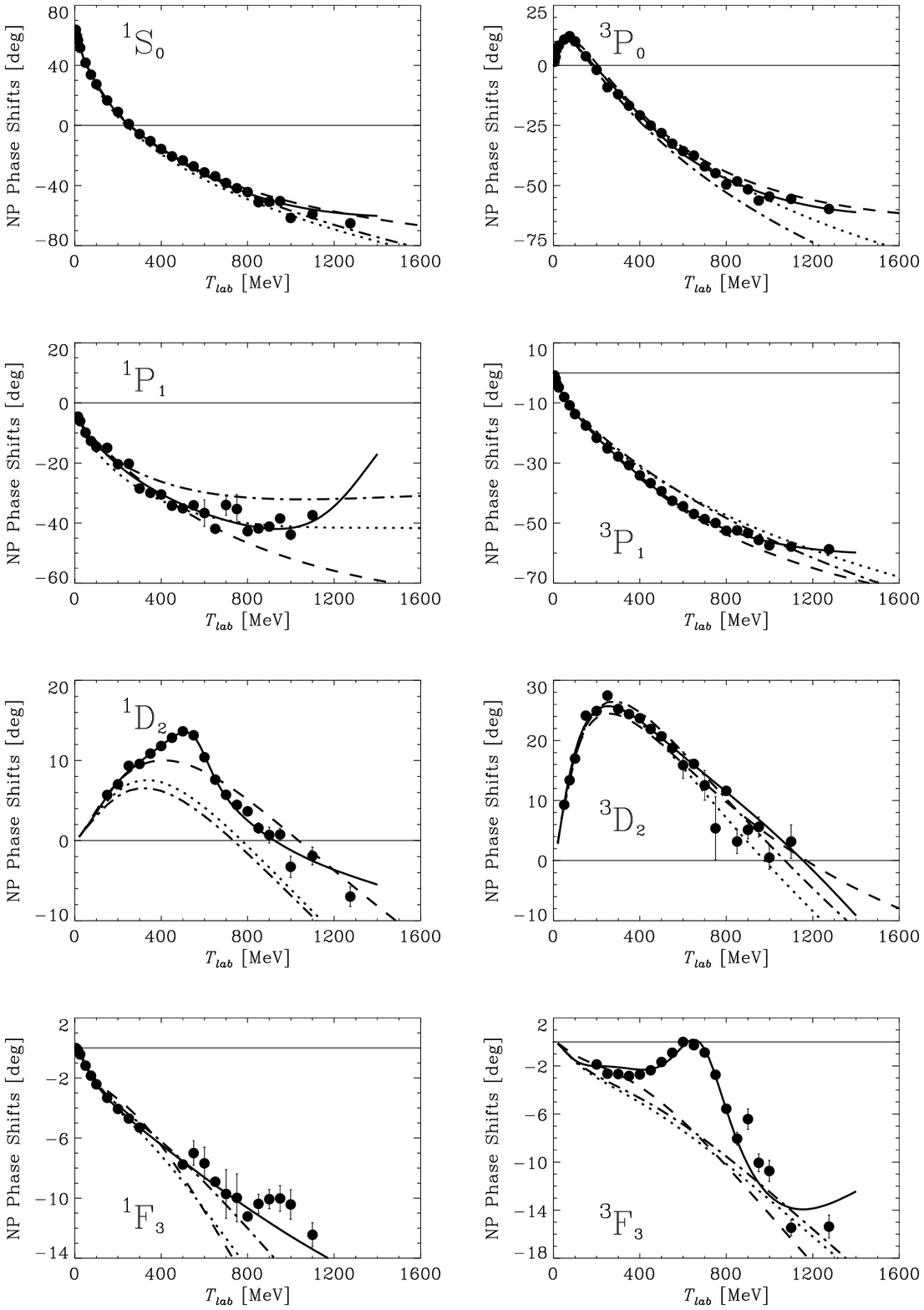,width=12cm}
\end{picture}
\caption[]{The phase shifts for
neutron-proton scattering in
the single channels.
The nomenclature is as in Fig.~\ref{sm97ppsc}.}
\label{sm97npsc}
\end{figure}
\begin{figure}[t]\centering
\begin{picture}(14,20)(0,-2)
\epsfig{file=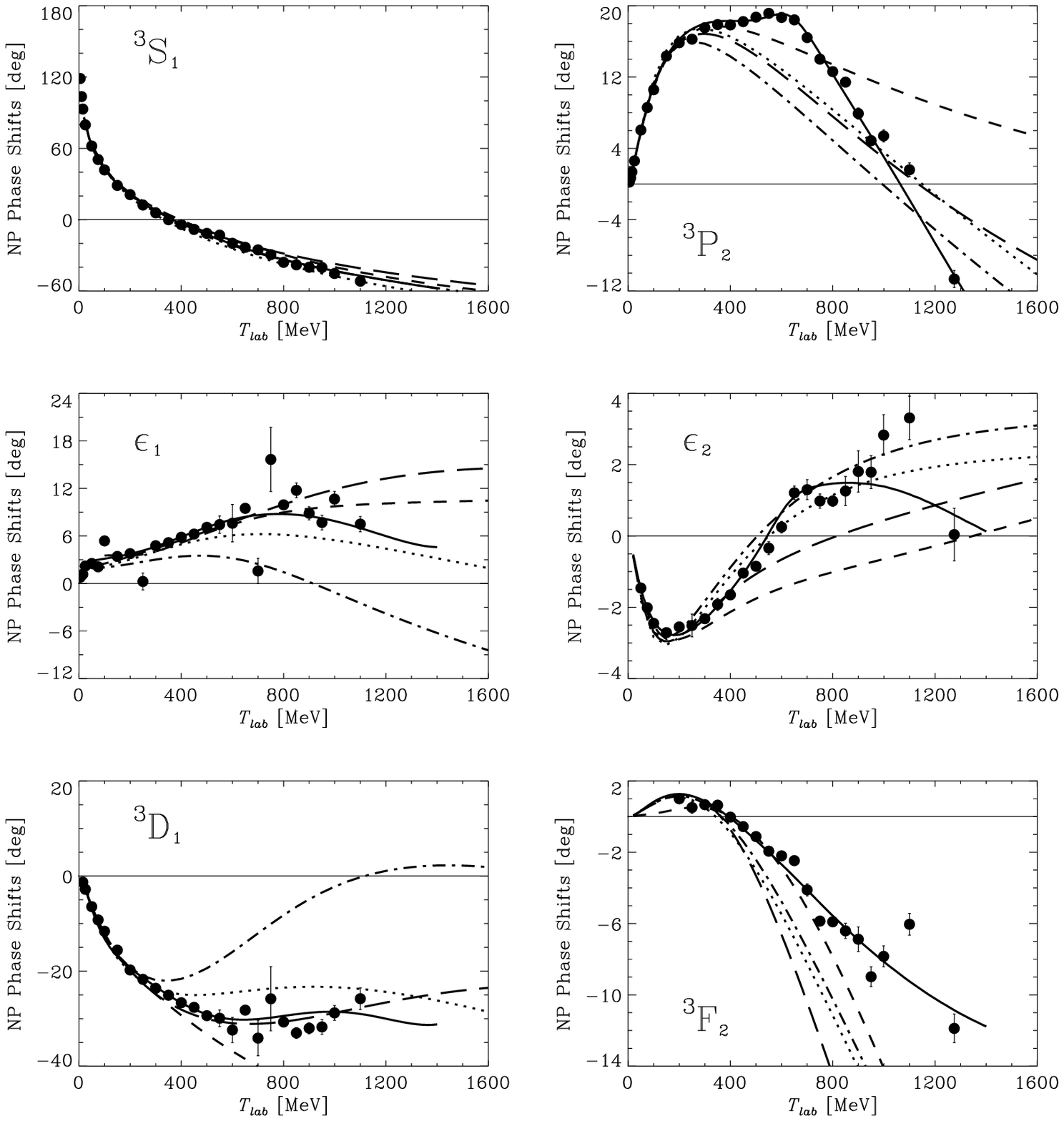,width=14cm}
\end{picture}
\caption[]{The phase shifts for
neutron-proton scattering in 
the ${}^3SD_1$ and ${}^3PF_2$ coupled channels.
The nomenclature is as in Fig.~\ref{sm97ppsc}.}
\label{sm97npcc}
\end{figure}
\begin{figure}[t]\centering
\begin{picture}(14,20)(0,0)
\epsfig{file=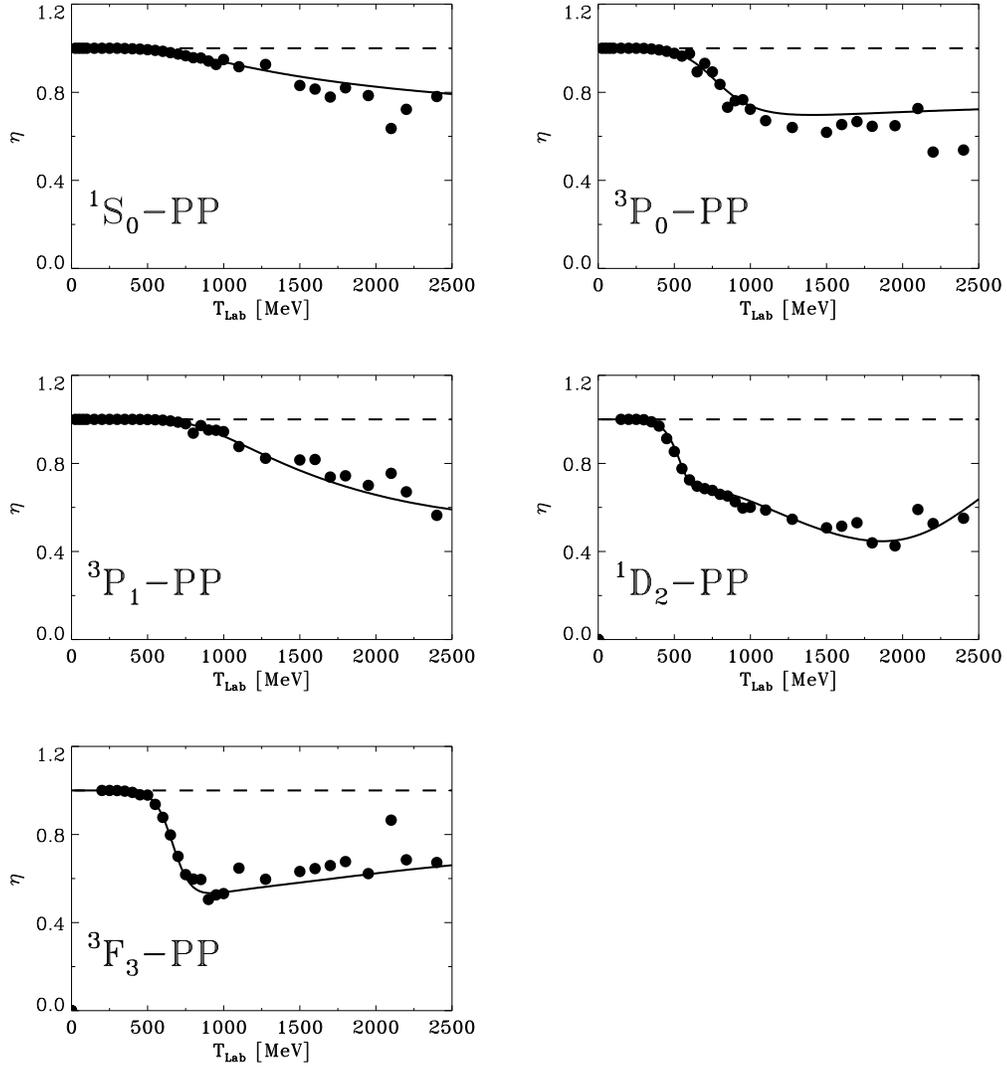,width=14cm}
\end{picture}
\caption[]{The absorption in 
proton-proton uncoupled channels.
The full dots are  SM97 single energy fits while
the solid curves represent the SM97 continuous
energy fits as well as the final results of our optical model
searches.}
\label{sm97ppeta}
\end{figure}
\begin{figure}[t]\centering
\begin{picture}(14,20)(0,0)
\epsfig{file=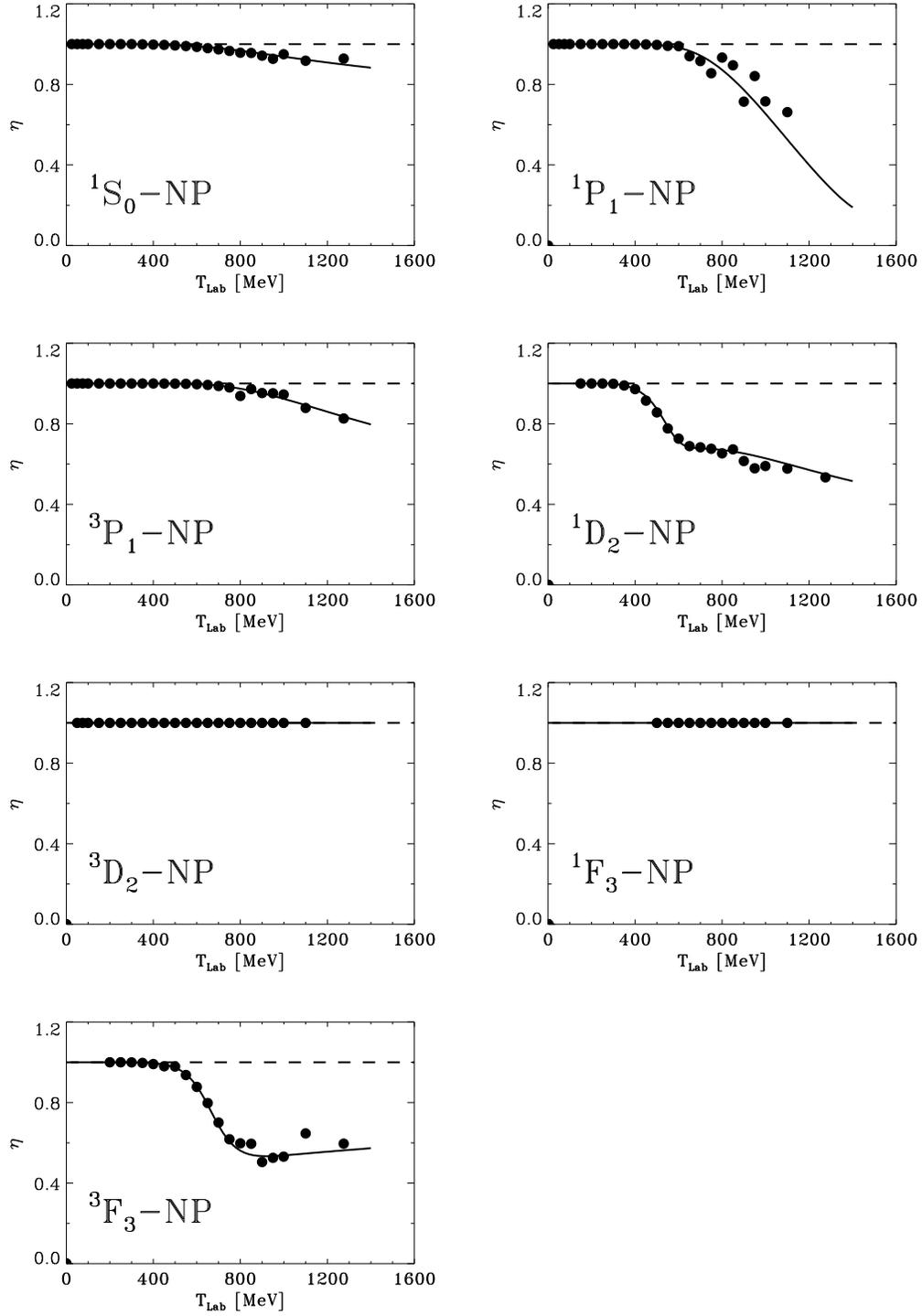,width=14cm}
\end{picture}
\caption[]{The absorption
in neutron-proton
uncoupled channels.
The nomenclature is as in Fig.~\ref{sm97ppeta}.}
\label{sm97npeta}
\end{figure}
\begin{figure}[t]\centering
\begin{picture}(14,20)(0,0)
\epsfig{file=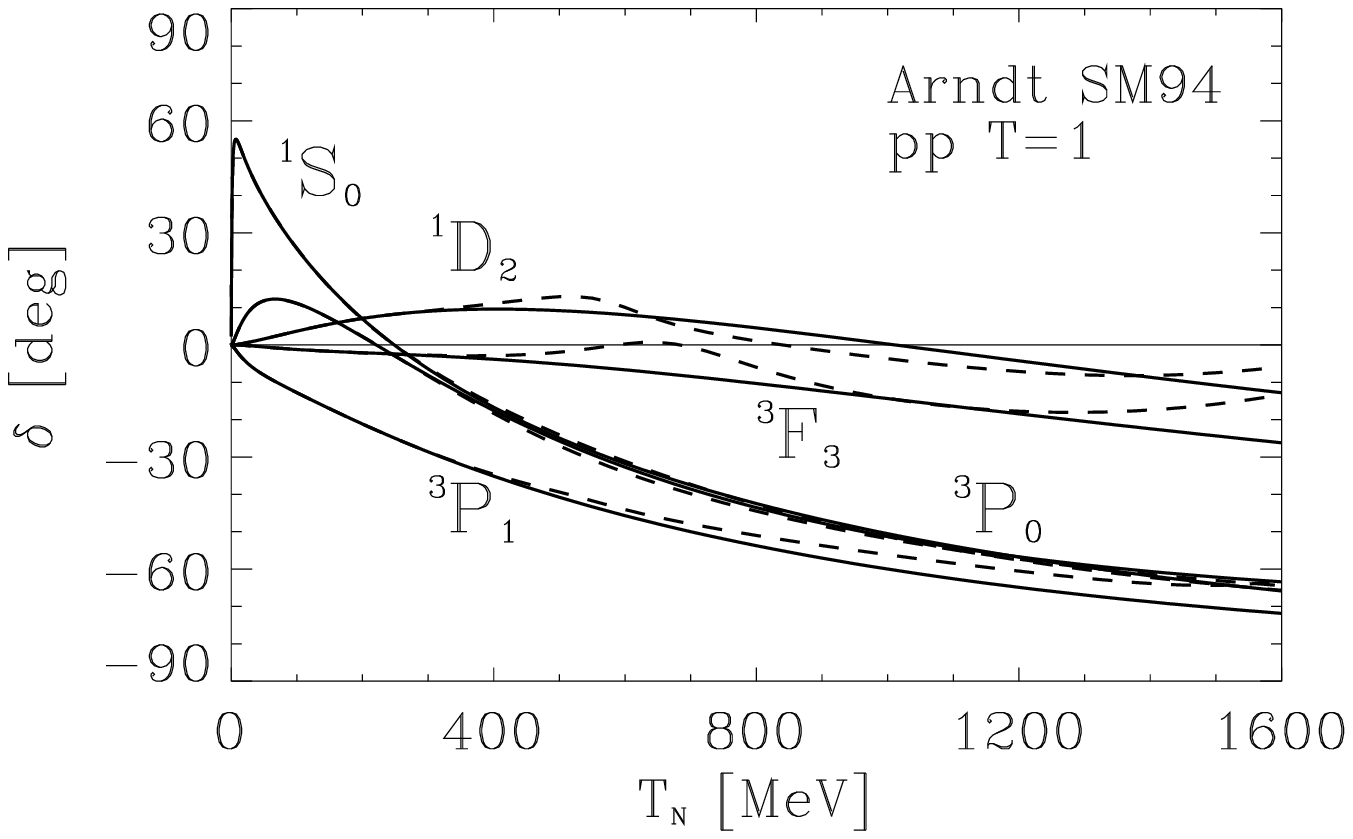,width=14cm}
\end{picture}
\caption[]{ The real 
phase shifts of SM94 for T= 1 proton-proton
scattering (dashed curves) compared to the phase shifts 
given from our inversion potentials (solid curves).
}
\label{sm94scppt1phase}
\end{figure}
\begin{figure}[t]\centering
\begin{picture}(14,20)(0,0)
\epsfig{file=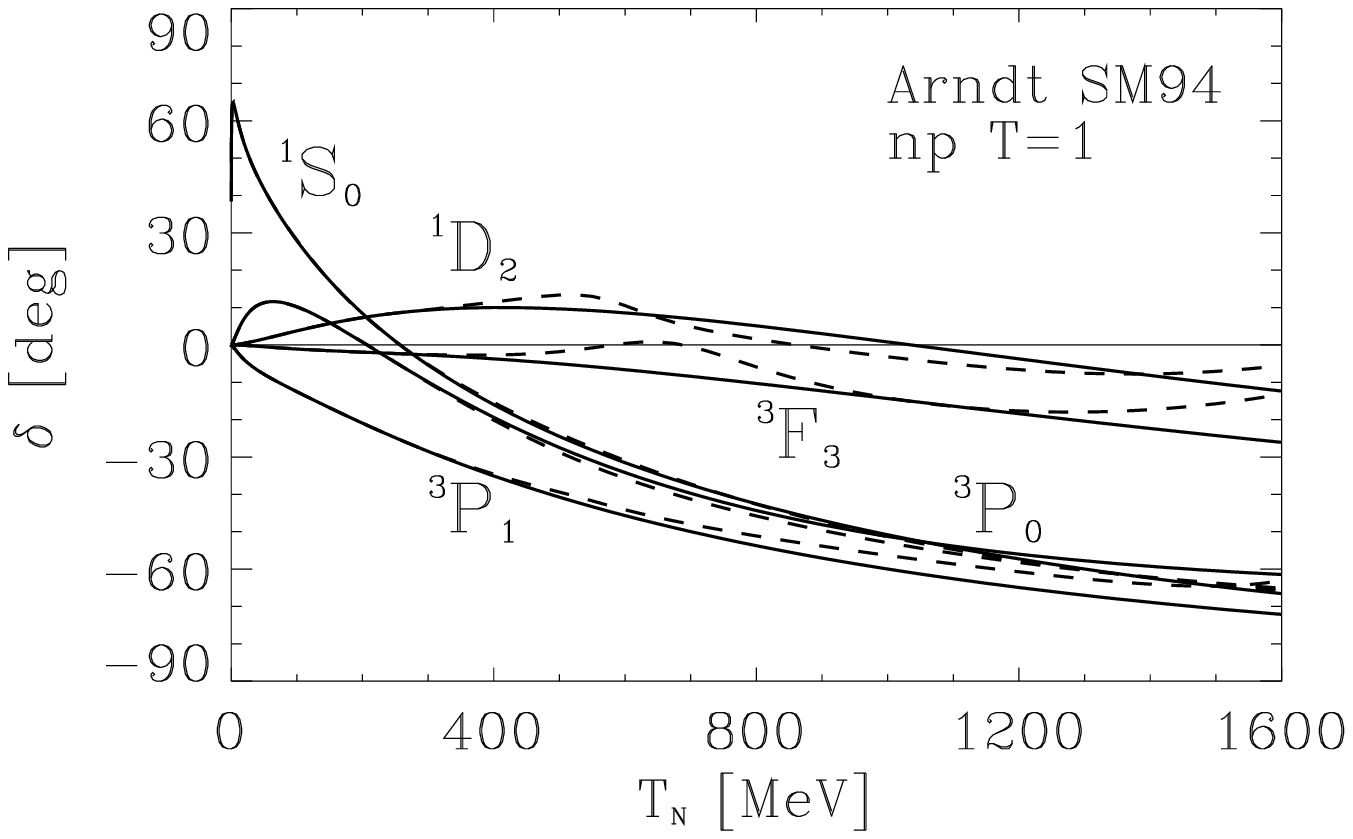,width=14cm}
\end{picture}
\caption[]{ The real 
phase shifts of SM94 for T= 1 neutron-proton
scattering (dashed curves) compared to the phase shifts 
given from our inversion potentials (solid curves).
}
\label{sm94scnpt1phase}
\end{figure}
\begin{figure}[t]\centering
\begin{picture}(14,20)(0,0)
\epsfig{file=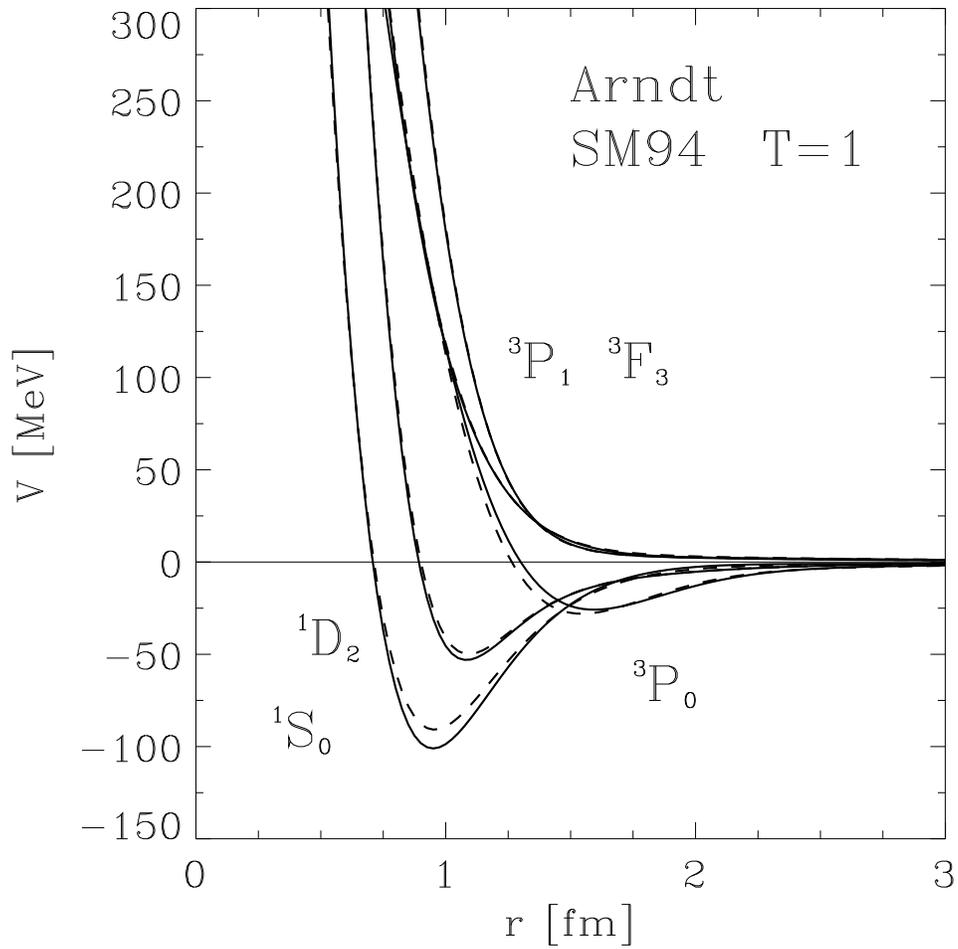,width=14cm}
\end{picture}
\caption[]{The 
inversion potentials from the SM94 
T = 1 uncoupled channel phase shifts.
The potentials from inversion of  neutron-proton 
 and proton-proton data are displayed by the solid and
 dashed curves respectively.} 
\label{sm94sct1pot}
\end{figure}
\begin{figure}[t]\centering
\begin{picture}(14,20)(0,0)
\epsfig{file=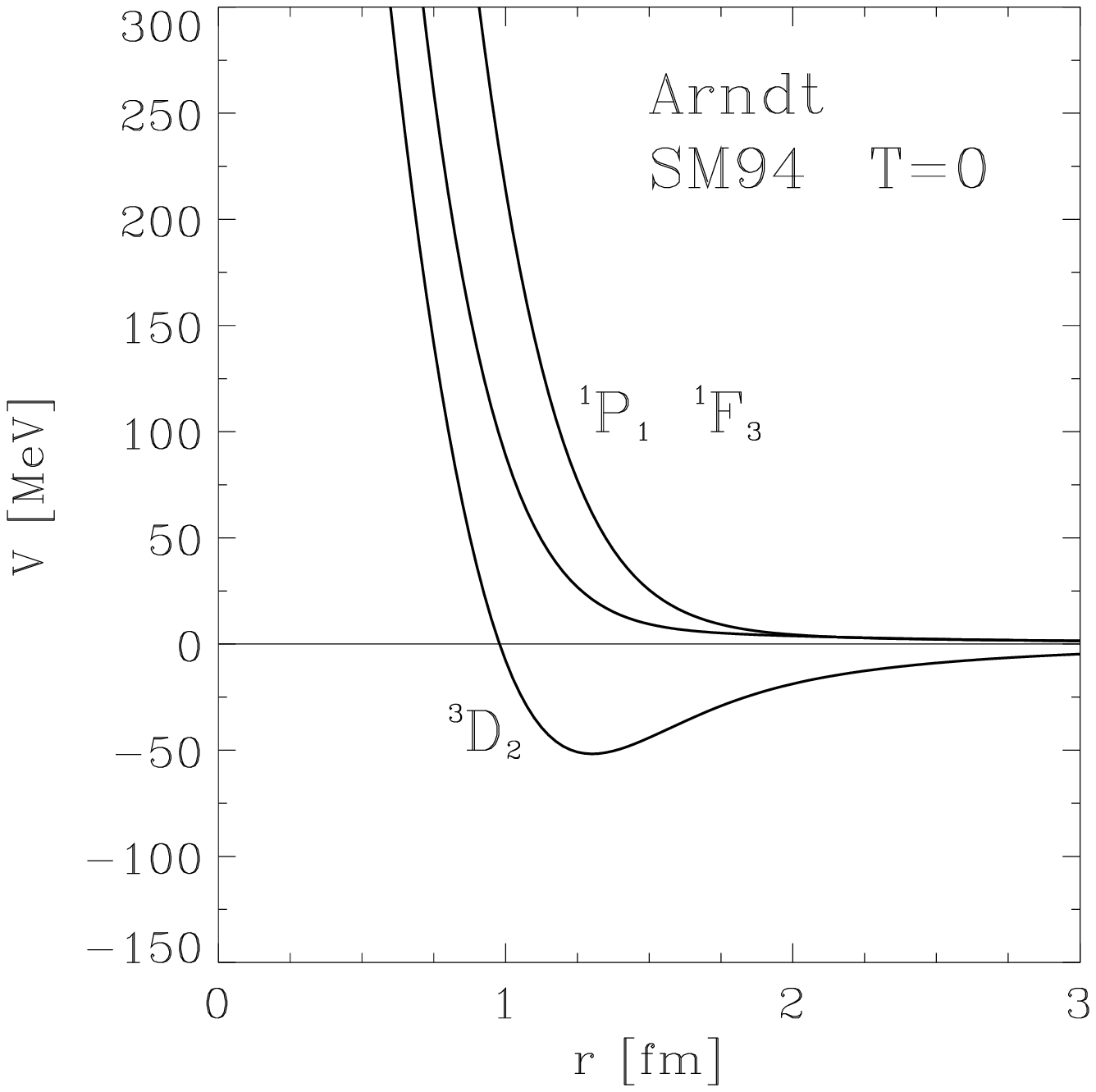,width=14cm}
\end{picture}
\caption[]{The 
inversion potentials from the SM94 T = 0
uncoupled channel phase shifts. 
The nomenclature is as in 
Fig.~\ref{sm94sct1pot}.}
\label{sm94sct0pot}
\end{figure}
\begin{figure}[t]\centering
\begin{picture}(7,10)(0,0)
\epsfig{file=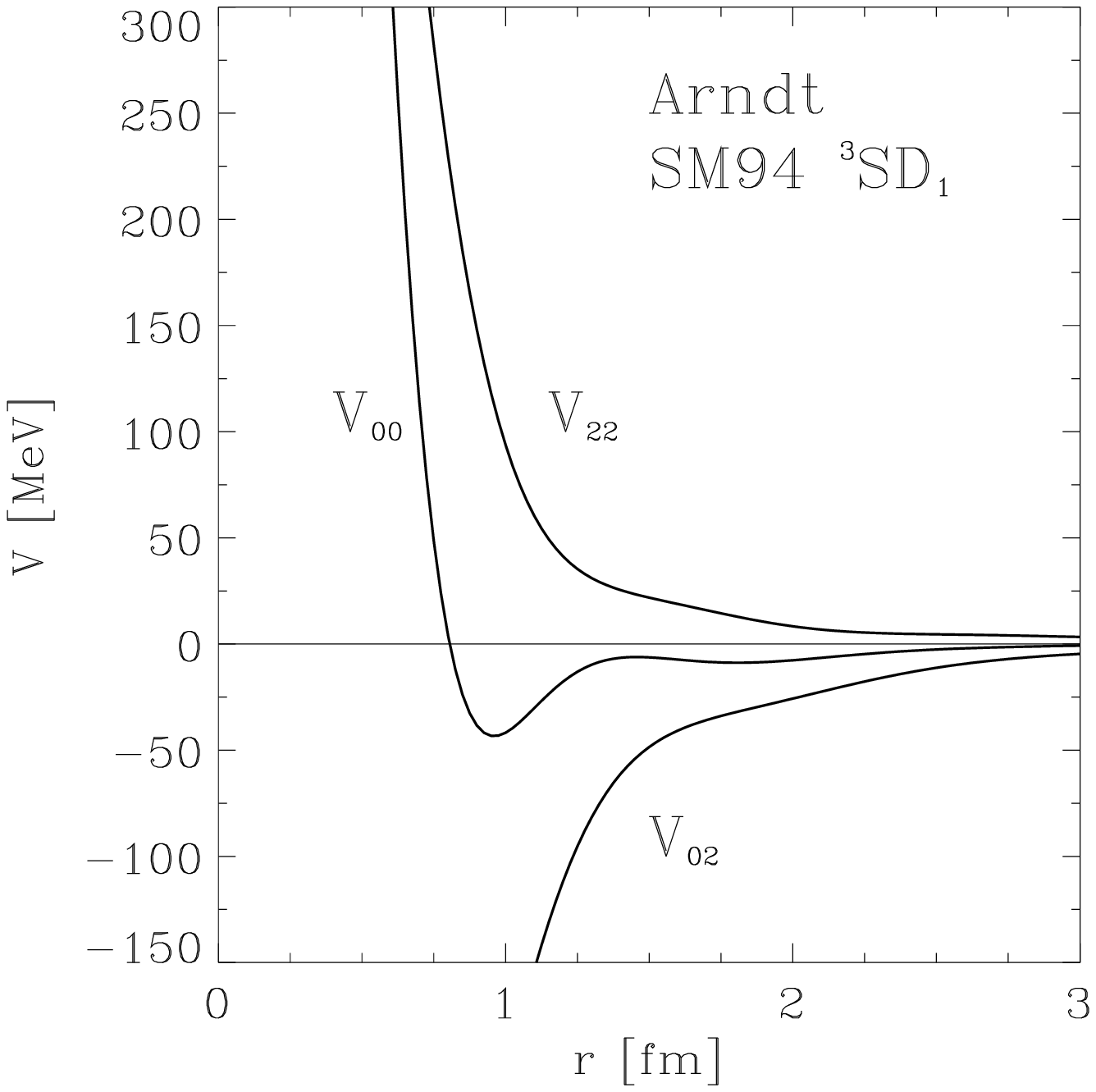,width=7cm}
\end{picture}
\begin{picture}(7,10)(0,0)
\epsfig{file=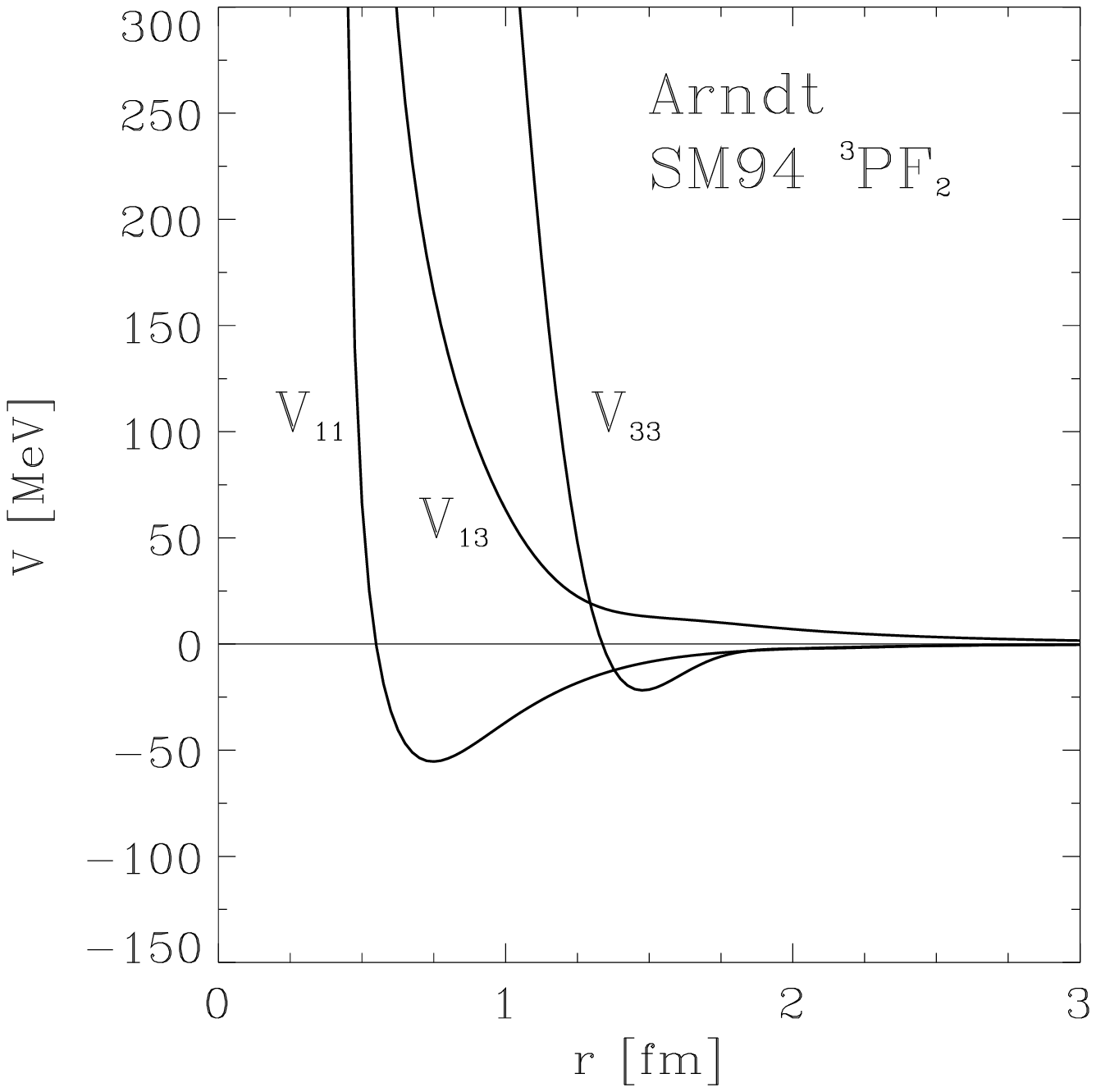,width=7cm}
\end{picture}
\caption[]{The 
inversion potentials from the SM94 coupled
channel phase shifts. 
The nomenclature is as in 
Fig.~\ref{sm94sct1pot}.}
\label{sm94cct0t1pot}
\end{figure}
\begin{figure}[t]\centering
\begin{picture}(14,20)(0,0)
\epsfig{file=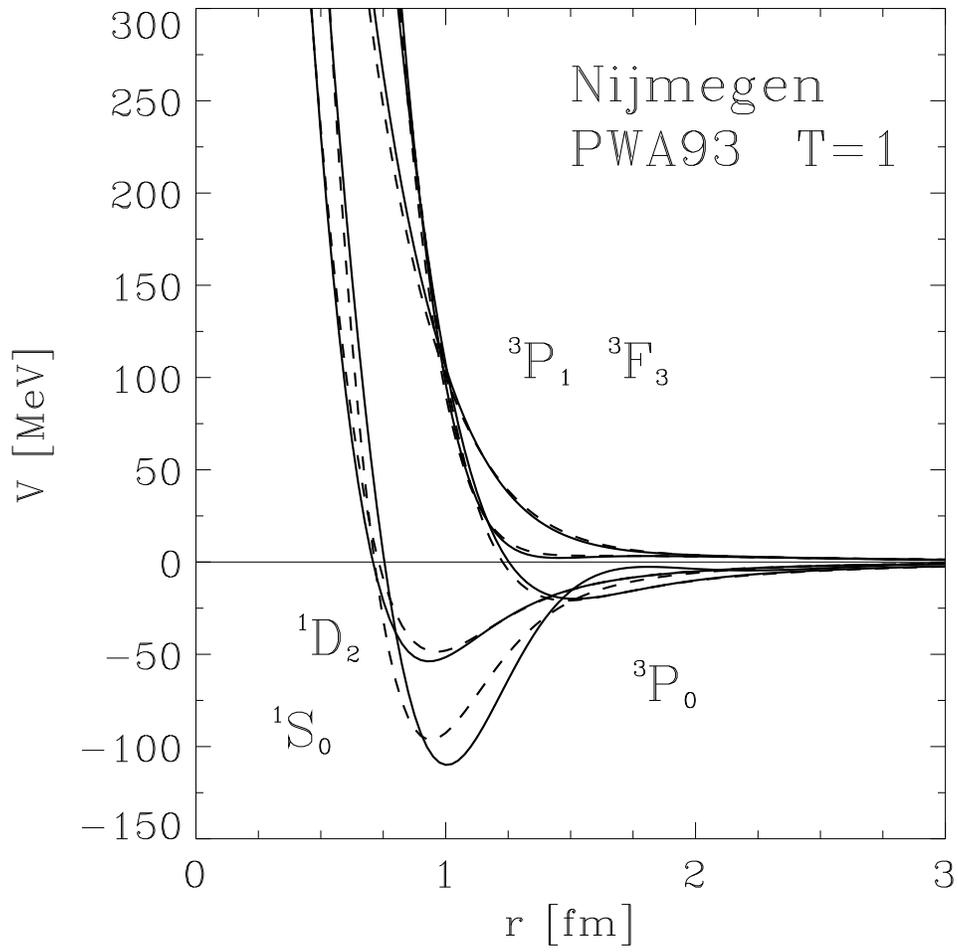,width=14cm}
\end{picture}
\caption[]{The 
inversion potentials from the Nijmegen PWA93 
T = 1 uncoupled channel phase shifts. 
The nomenclature is as in 
Fig.~\ref{sm94sct1pot}.}
\label{pwa93scpot}
\end{figure}
\begin{figure}[t]\centering
\begin{picture}(14,20)(0,0)
\epsfig{file=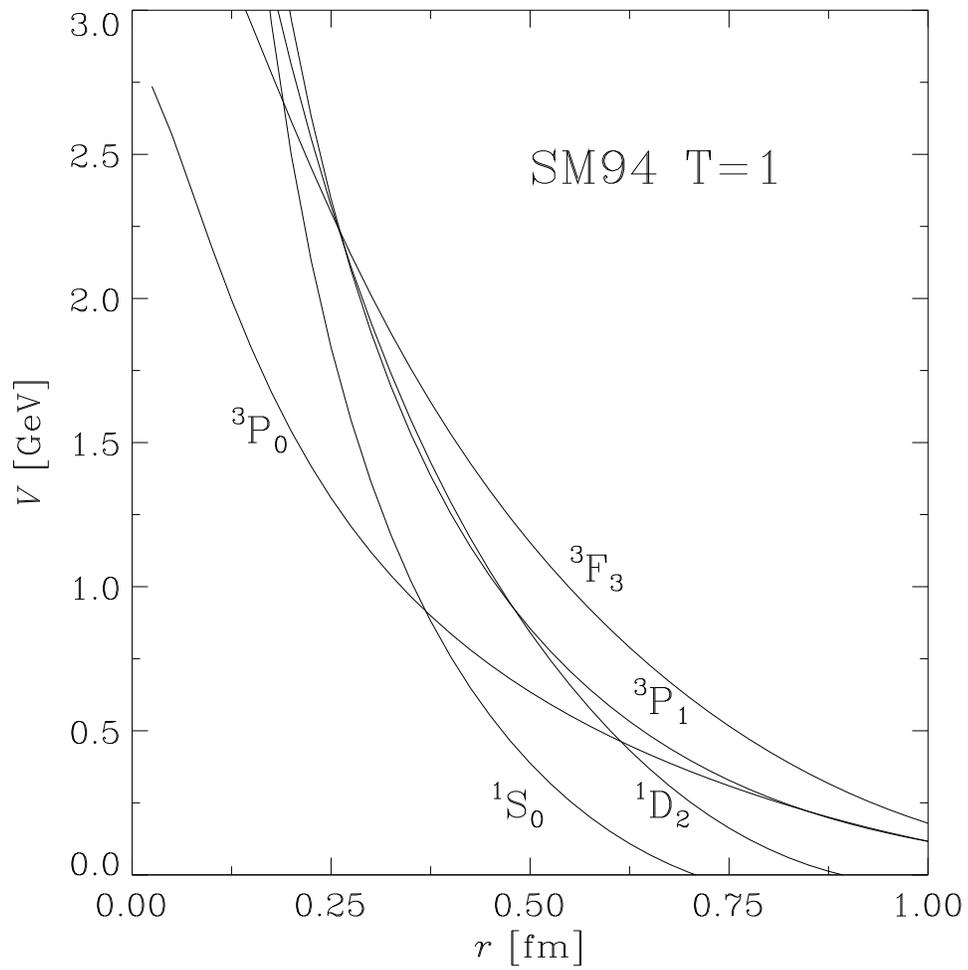,width=14cm}
\end{picture}
\caption[]{The short range properties of
the proton-proton
inversion potentials from the T = 1 SM94 phase shifts.}
\label{sm94t1core}
\end{figure}
\begin{figure}[t]\centering
\begin{picture}(14,20)(0,0)
\epsfig{file=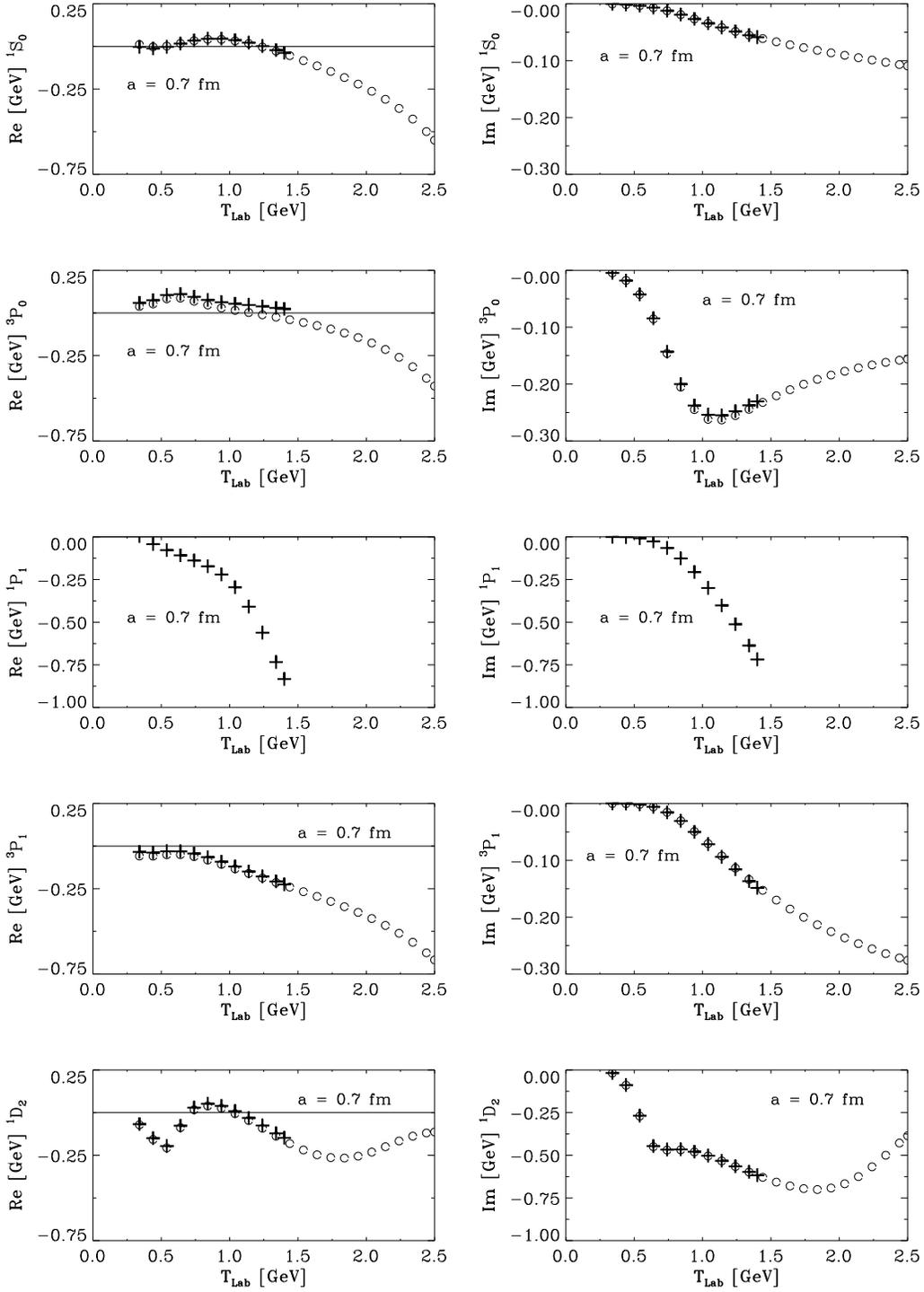,width=14cm}
\end{picture}
\caption[]{The optical model potential
strengths as  functions of energy
for Gaussian forms with range 0.7 fm
for uncoupled L=0--2 channels.
The real strengths, $V_0(T_{\rm Lab})$, are shown on the left
and the imaginary ones, $W_0(T_{\rm Lab})$,
on the right.
The circles and crosses
depict  the results of our analyses of the $pp$ data
to 2.5 GeV and 
of the SM97 $np$ data respectively.}
\label{07allsc}
\end{figure}
\begin{figure}[t]\centering
\begin{picture}(14,20)(0,0)
\epsfig{file=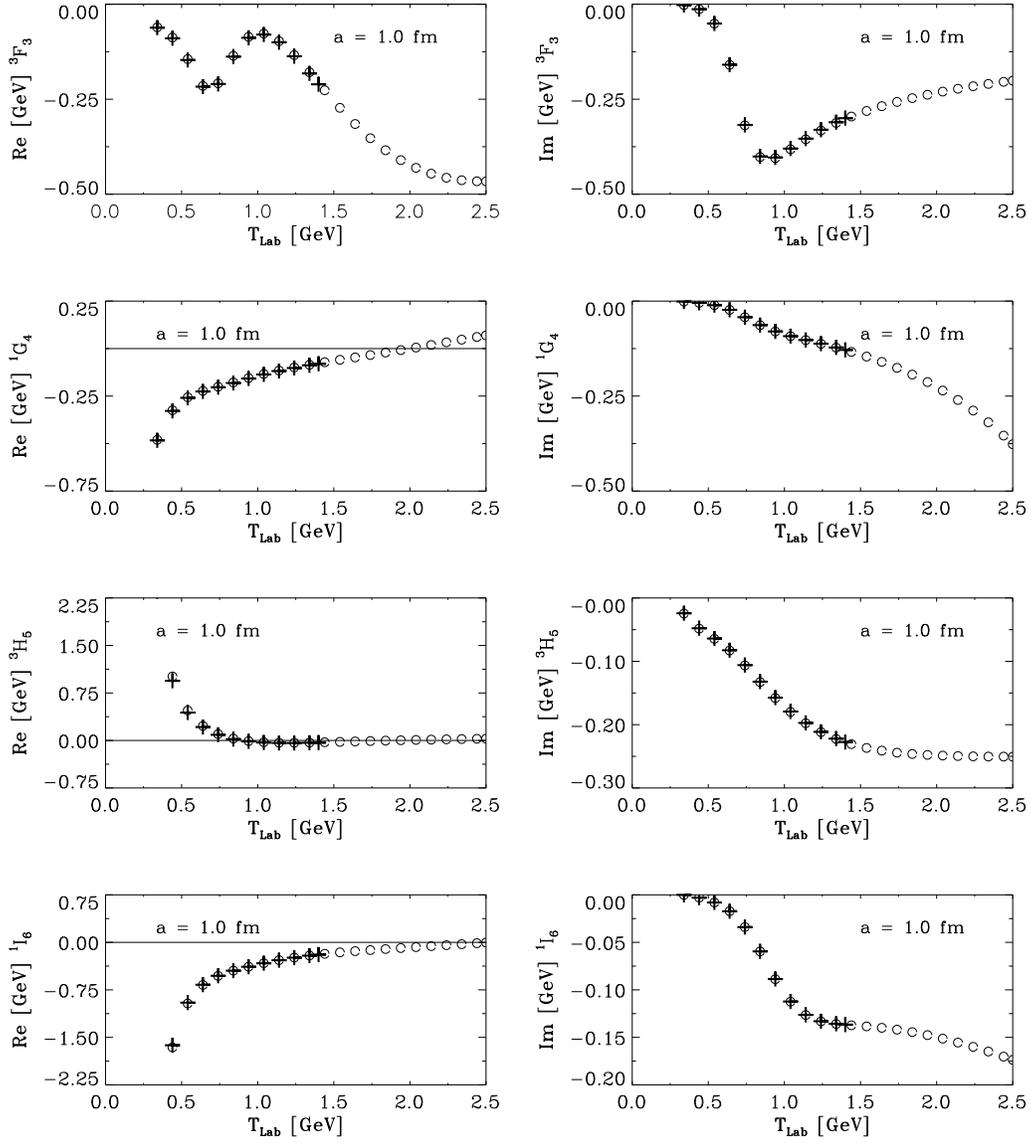,width=14cm}
\end{picture}
\caption[]{As for Fig.~\ref{07allsc}
but for the uncoupled L=3--6 channels
and with a Gaussian range of 1.0 fm.
}
\label{10allsc}
\end{figure}
\begin{figure}[t]\centering
\begin{picture}(14,20)(0,0)
\epsfig{file=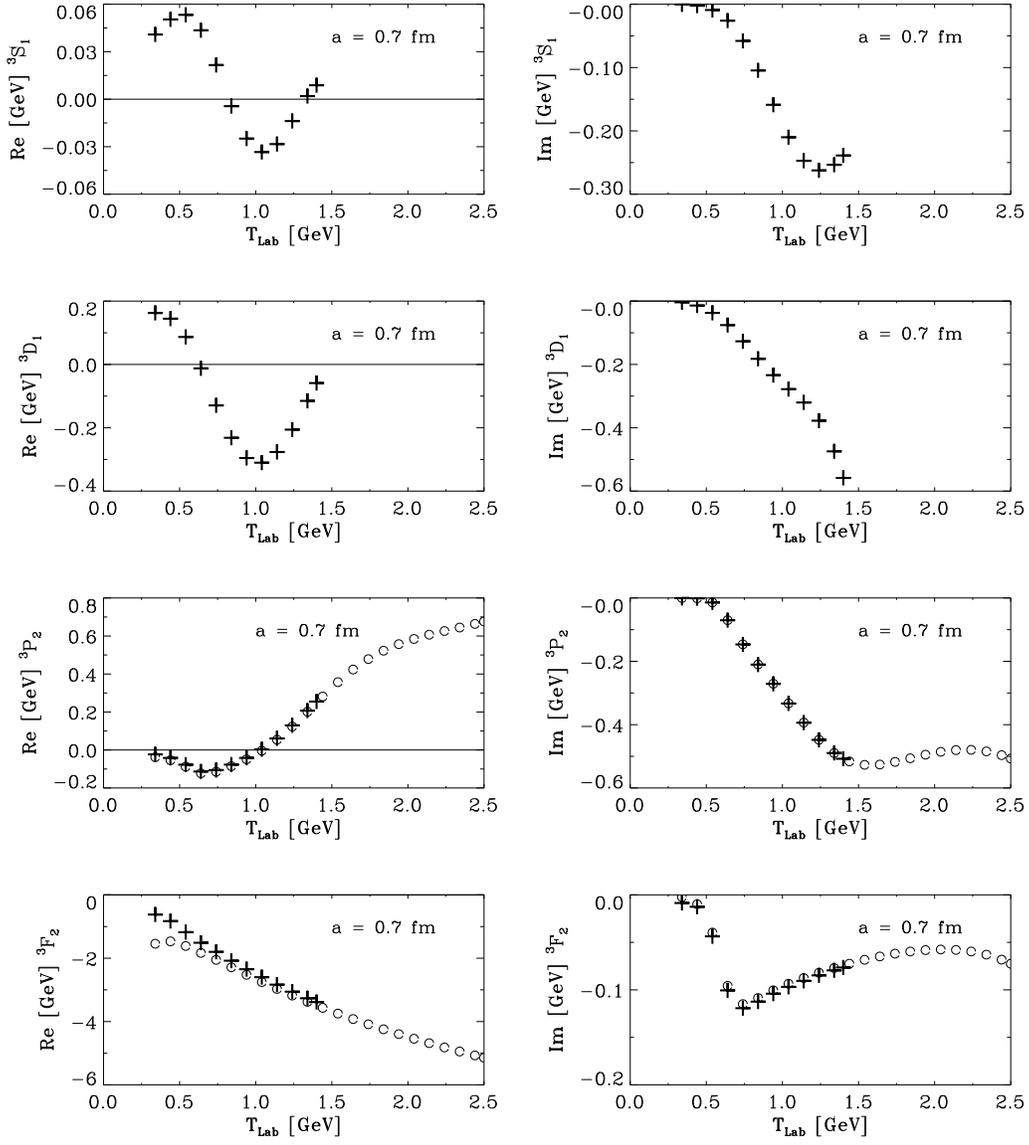,width=14cm}
\end{picture}
\caption[]{As for Fig.~\ref{07allsc}
but for the coupled channels ${}^3SD_1$ and ${}^3PF_2$.}
\label{07allcc}
\end{figure}
\begin{figure}[t]\centering
\begin{picture}(21,21)(2,3)
\epsfig{figure=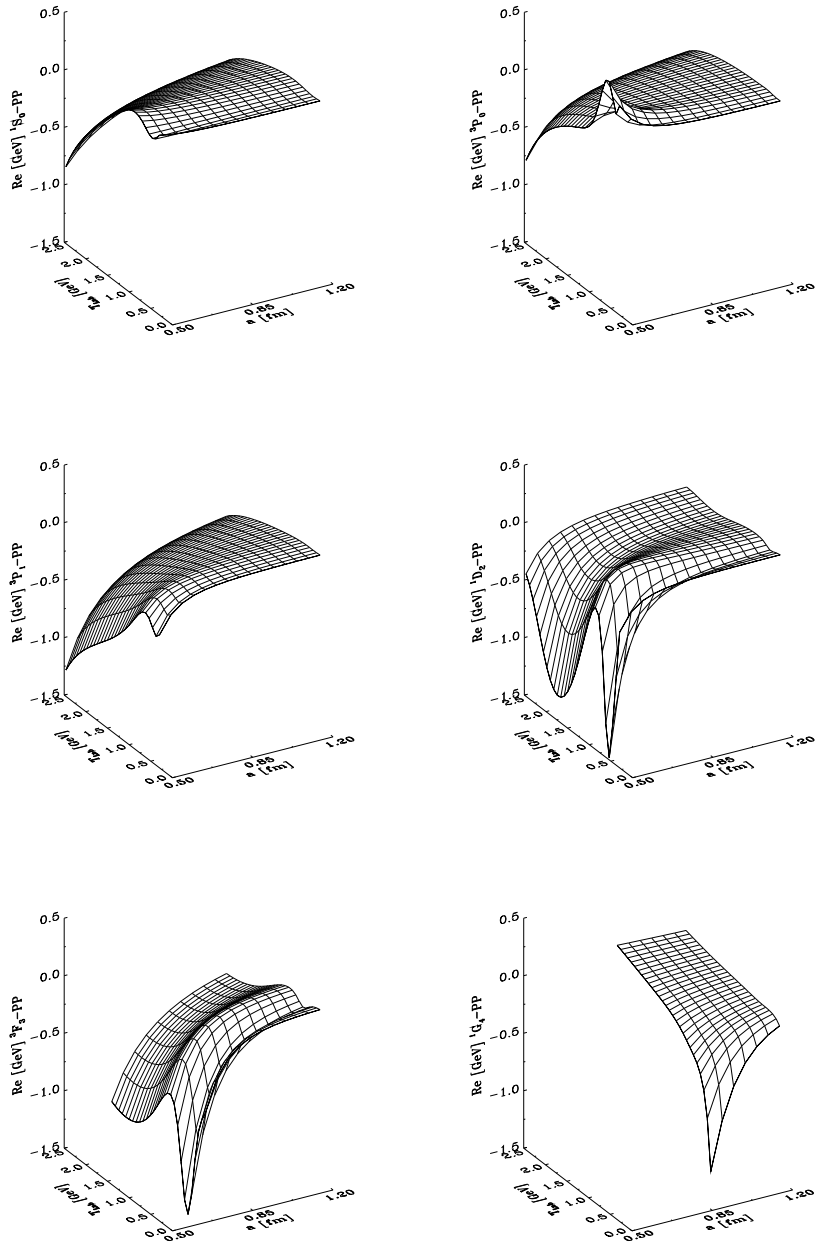,width=21cm}
\end{picture}
\caption[]{The variations of the real parts of the 
$pp$ optical potentials in 
various channels as functions of
$T_{\rm Lab}$ and of the range of
 the Gaussian form.}
\label{potre}
\end{figure}
\begin{figure}[t]\centering
\begin{picture}(21,21)(2,3)
\epsfig{file=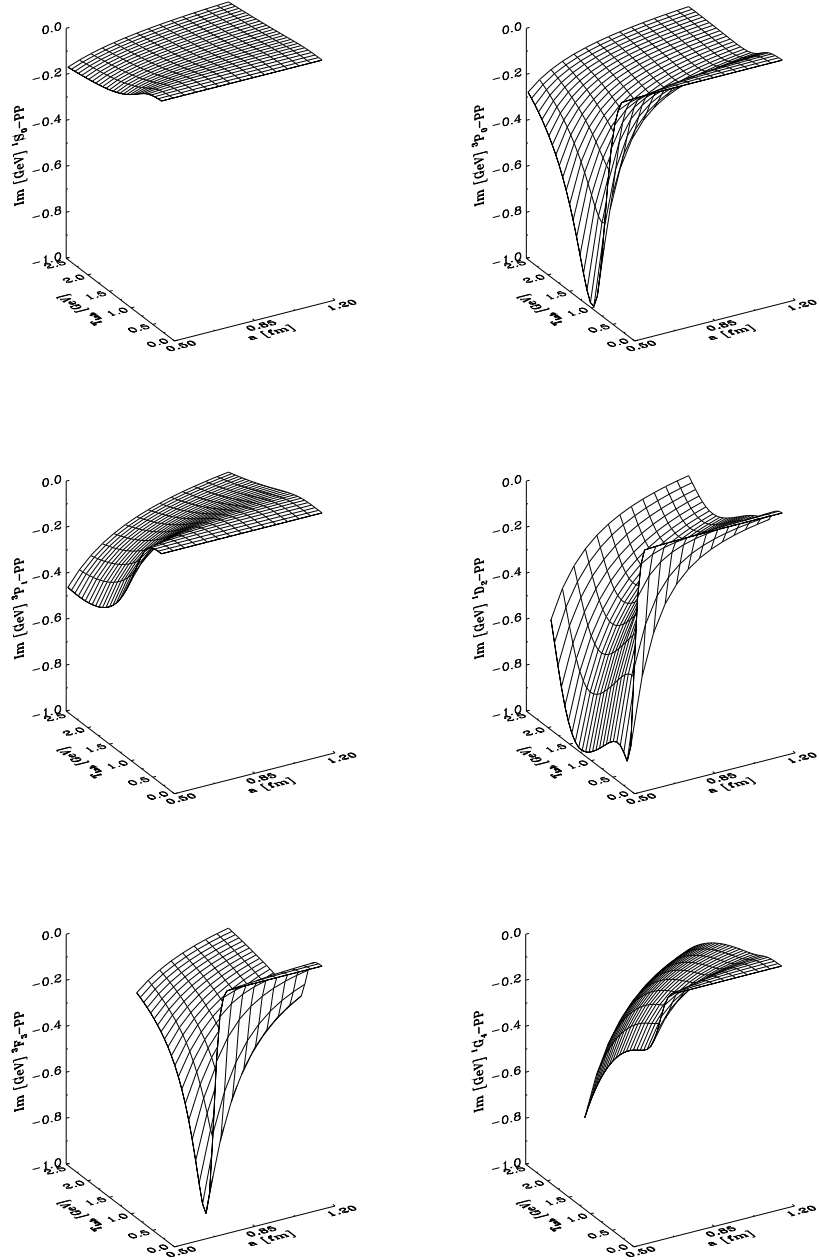,width=21cm}
\end{picture}
\caption[]{As for Fig.~\ref{potre}
but for the imaginary potentials.}
\label{potim}
\end{figure}
\begin{figure}[t]\centering
\begin{picture}(14,20)(0,0)
\epsfig{file=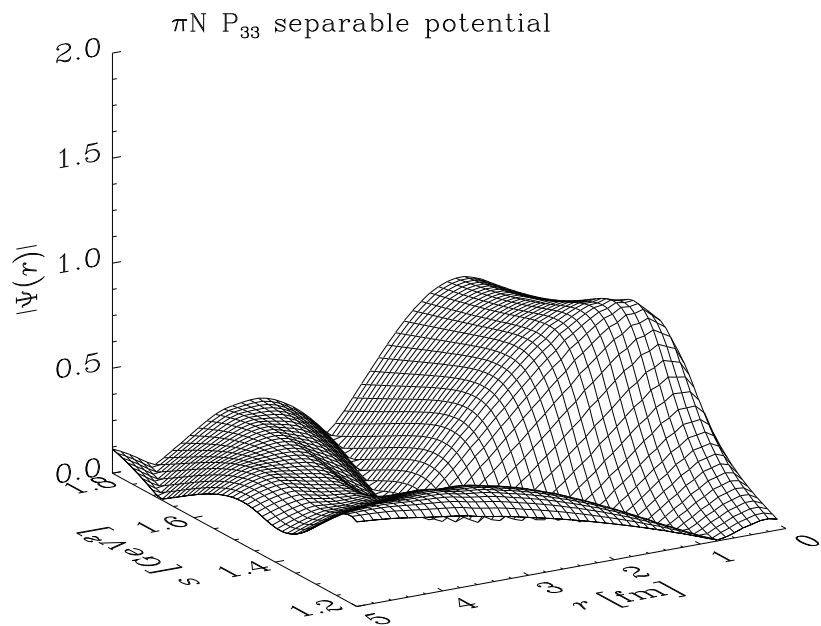,width=14cm}
\end{picture}
\caption[]{The modulus of the $\pi N$ wave function 
in the $P_{33}$ channel of the separable
interaction defined in the text in coordinate space
and as a function of $s$.}
\label{sepwf}
\end{figure}
\begin{figure}[t]\centering
\begin{picture}(14,20)(0,0)
\epsfig{file=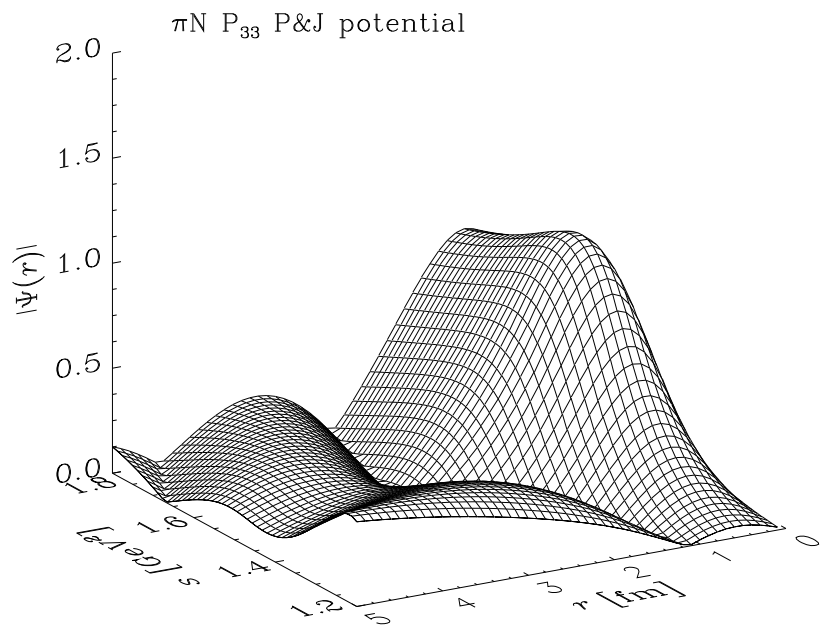,width=14cm}
\end{picture}
\caption[]{As for Fig.~\ref{sepwf}
except for the Pearce-Jennings interaction.}
\label{pjwf}
\end{figure}
\begin{figure}[t]\centering
\begin{picture}(14,20)(0,0)
\epsfig{file=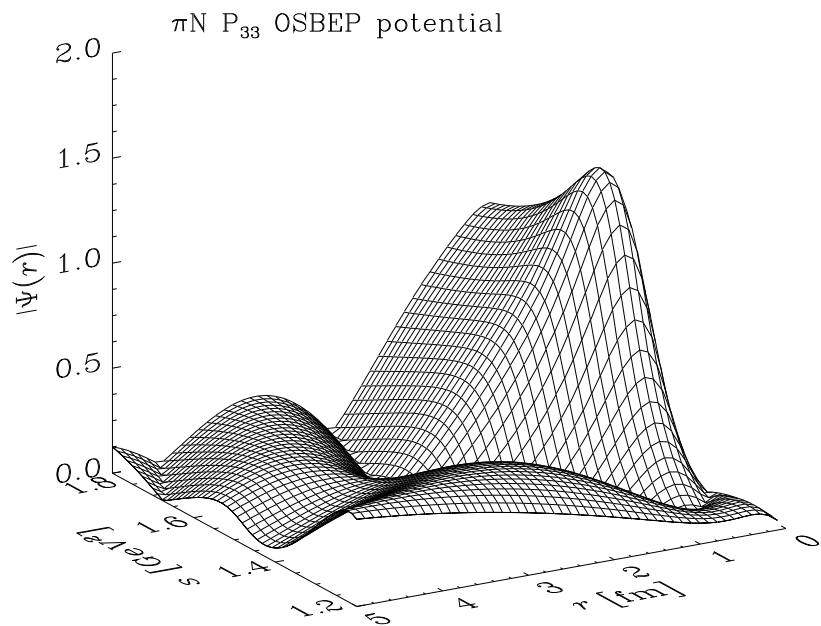,width=14cm}
\end{picture}
\caption[]{As for Fig.~\ref{sepwf}
except for the OSBEP interaction.}
\label{osbwf}
\end{figure}
\begin{figure}[t]\centering
\begin{picture}(14,20)(0,0)
\epsfig{file=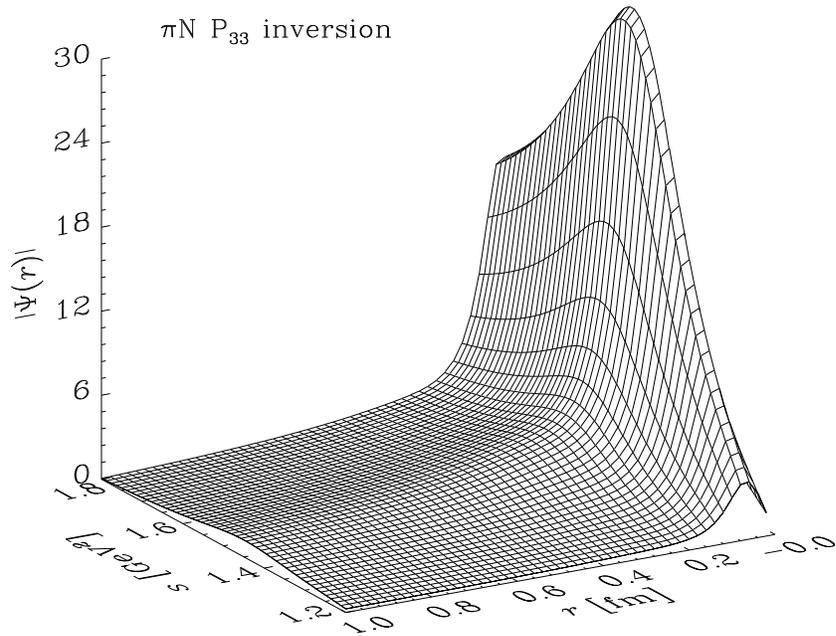,width=14cm}
\end{picture}
\caption[]{As for Fig.~\ref{sepwf}
except for the inversion interaction.}
\label{invwf}
\end{figure}
\begin{figure}[t]\centering
\begin{picture}(14,20)(0,0)
\epsfig{file=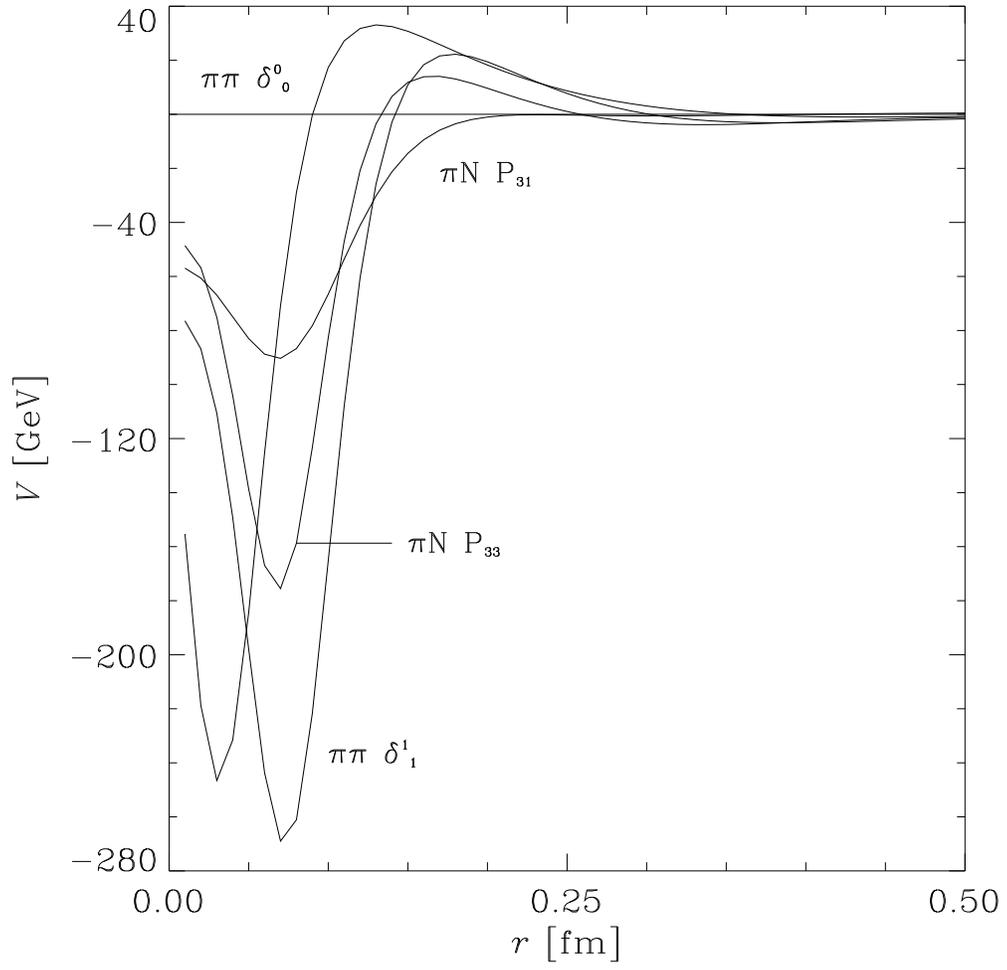,width=14cm}
\end{picture}
\caption[]{The short range potentials found by inversion
of $\pi N$ and  $\pi \pi$ phase shift data.}
\label{hadronpot}
\end{figure}

\end{document}